\documentclass[12pt, twocolumn,showpacs,preprintnumbers,aps,prl,reprint]{revtex4-1}

\usepackage[pdftex]{graphicx}
\usepackage{subfigure}
\usepackage{color}
\usepackage{amsmath}
\usepackage[linktocpage,colorlinks=true,linkcolor=blue,citecolor=blue,breaklinks=true]{hyperref}
\usepackage{verbatim}
\usepackage{breakcites}

\definecolor{cream}{RGB}{222,217,201}
\definecolor{red}{RGB}{255,0,0}

\newcommand{\xiI}{\xi_{\mathrm{I}}}
\newcommand{\thetaI}{\theta_{\mathrm{I}}}
\newcommand{\thetaS}{\theta_{\mathrm{S}}}

\newcommand{\K}{\mathsf{K}_{\cos\thetaI}}
\newcommand{\F}{\mathsf{F}_{\cos\thetaI}}
\newcommand{\sn}{\mathrm{sn}_{\cos\thetaI}}

\begin{document}

\preprint{}

\title{Motility of active nematic films driven by ``active anchoring''}

\author{Matthew L. Blow}
\author{Marco Aqil}
\author{Benno Liebchen}
\author{Davide Marenduzzo}
\affiliation{SUPA, School of Physics and Astronomy, University of Edinburgh, James Clerk Maxwell Building, Peter Gutherie Tait Road, Edinburgh EH9 3FD, UK.}
\email[]{E-mail: dmarendu@ph.ed.ac.uk}

\date{\today}

\begin{abstract}
We provide a minimal model for an active nematic film in contact with both a solid substrate and a passive isotropic fluid, and explore its dynamics in one and two dimensions using a combination of hybrid Lattice Boltzmann simulations and analytical calculations. By imposing nematic anchoring at the substrate while active flows induce a preferred alignment at the interface (``active anchoring''), we demonstrate that directed fluid flow spontaneously emerges in cases where the two anchoring types are opposing. In one dimension, our model reduces to an analogue of a loaded elastic column. Here, the transition from a stationary to a motile state is akin to the buckling bifurcation, but offers the possilibity to reverse the flow direction for a given set of parameters and boundary conditions solely by changing initial conditions. The two-dimensional variant of our model allows for additional tangential instabilities, leading to self-assembled propagating surface waves for intermediate activity and for a continously deforming irregular surface at high activity. Our results might be relevant for designing active microfluidic geometries, but also for curvature-guided self-assembly or switchable diffraction gratings.
\end{abstract}
%%%END OF ABSTRACT%%%%

\pacs{}

\maketitle

\section{Introduction}
\label{sec:intro}
``Active matter'' refers to systems composed of interacting particles that generate motion via the local injection of energy. A diverse range of biological and synthetic systems, including microtubule bundles~\cite{Sanchez2012}, actin filaments~\cite{Prost2015}, suspensions of swimming bacteria~\cite{Dunkel2013} and chemically propelled colloids~\cite{Golestanian2005}, fall under this designation and show a remarkable universality. Being non-equilibrium systems, such active materials exhibit many novel properties including collective motion, turbulent-like pattern formation and the proliferation of topological defects~\cite{Giomi2012,Giomi2013,Thampi2013,Giomi2014,Tjhung2015}.

An area of intense interest is active matter in context with surfaces or interfaces, such as active droplets~\cite{Joanny2012,Tjhung2012,GiomiDeSimone,Tjhung2015,Khoromskaia2015} and films~\cite{Voituriez2005,Sankararaman2009}. Such systems have been found to self-organise into states where they undergo spontaneous fluid motion, which can be transmitted to the wider system. For instance, cytoplasmic streaming, an important transport mechanism in cells, is powered by myosin motors walking on actin filaments at the cell boundaries~\cite{Goldstein2008,Woodhouse2013}. Active film equations have proven to form a useful framework for studying bacterial colonies and cellular monolayers~\cite{Doostmohammadi2015}, and active droplets provide a simple model of crawling cells~\cite{Tjhung2015}. The ability to produce controllable flow in these systems is also important in microfludics applications~\cite{KirbyBook}.

The interactions of the active film or drop with both the solid substrate and the surrounding inert fluid play a crucial role in determining its qualitative and quantitative behaviour. In models considered thus far (see {\it e.g.}~\cite{Voituriez2005,Davide2007}), the active fluid was taken to be strongly anchored at both interfaces -- that is, the nematic orientation was fixed relative to the interface by virtue of thermodynamic interactions, unrelated to activity.

But thermodynamics is not the only route to anchoring; ``active anchoring'' is a phenomenon by which activity-induced flows produce a preferred alignment at the interface between an active nematic and a passive isotropic fluid~\cite{Blow2014}. The resulting alignment is planar (director tangential to the interface) in active nematics where the constituent particles produce extensile stresses, and homeotropic (director perpendicular to the interface) for contractile active stresses.

Here, we provide a minimal model to explore the basic phenomenology of active nematics in contact with both a substrate and a fluid, where thermodynamic and active anchoring coexist and dictate the orientation at two different boundaries. In particular, our model describes a film of active nematic fluid that has strong thermodynamic anchoring to the solid substrate, but where the orientation at the interface with the surrounding fluid is determined solely by active anchoring. For simplicity, we neglect the aligning interaction between shear flow and the director; {\it i.e.} we consider a purely flow-tumbling material. This simplifies the analytical calculations and is useful here because the active anchoring discussed in~\cite{Blow2014} has been best characterised in this limit.

We find that, if the activity type is such that it produces active anchoring of the opposite type to the thermodynamic anchoring at the substrate (for example, extensile activity with homeotropic substrate anchoring, or contractile activity with planar substrate anchoring), then the nematic orientation may vary between the two interfaces in order to accommodate the conflicting anchoring conditions. The resulting nonuniformity in the active stress produces an active force, which drives a sustained flow in the film. 

We quantitatively study this phenomenon using numerical simulations and analytical calculations. To estabilish the basic principles by which active anchoring can cause spontaneous flow, we first consider a one-dimensional model in which the thickness of the film is spatially uniform and the nematic director varies only in the direction perpendicular to the substate. We find that the director profile and resulting flow are determined by a quantity $h'$ -- the non-dimensional film thickness in terms of an active lengthscale that accounts for the balance between activity and nematic elasticity. The governing equations are analogous to those for the much-studied engineering problem of a load-bearing elastic column~\cite{LoveBook}.

In the case of homeotropic anchoring with extensile activity, we identify a transition from a phase in which the director profile is uniform and the film stationary, to a state in which the profile is nonuniform and the film flows. This occurs when $h'$ surpasses a critical value, akin to the buckling transition for an elastic rod and similar to the flow transition seen when there is thermodynamic anchoring at both interfaces~\cite{Voituriez2005}. 
When the substrate anchoring is not perpendicular to the substrate, the symmetry of flow-direction is broken. Depending on the initial conditions, we observe a ``positive'' state that permits non-zero flow for all non-zero $h'$, and, remarkably, also a ``negative'' state in a counterintuitive direction, which is suppressed below a certain $h'$-threshold.

Beyond these 1D effects, tangential instabilities in active films can produce undulations or even cause film break-up. 
To demonstrate these effects, we also consider a two-dimensional model that additionally permits variation of the film thickness and the nematic director tangentially to the substrate. For sufficiently low $h'$ the film remains uniform in the tangential direction, even when undergoing spontaneous flow, but larger values of $h'$ lead to surface instability and the development of undulations at the interface. The amplitude of these surface undulations stabilises at a constant value leading to regular surface waves, which may travel at a different speed to the underlying fluid. For even larger $h'$ values, the film undergoes irregular and non-steady deformations. In some cases it breaks up and ejects ``blobs'' of active material into the isotropic phase, sometimes leading to the creation of topological defects. Since the surface tension of the interface resists such deformations, the thresholds of $h'$ at which they occur increases with surface tension.

\section{Model}
\label{sec:system}
\begin{figure*}[htp]
\begin{center}
{\includegraphics[height=4cm]{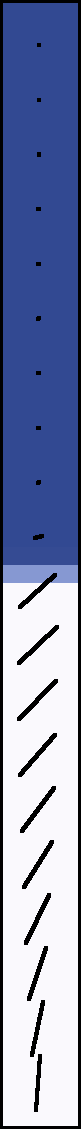}
}
{\includegraphics[height=4cm]{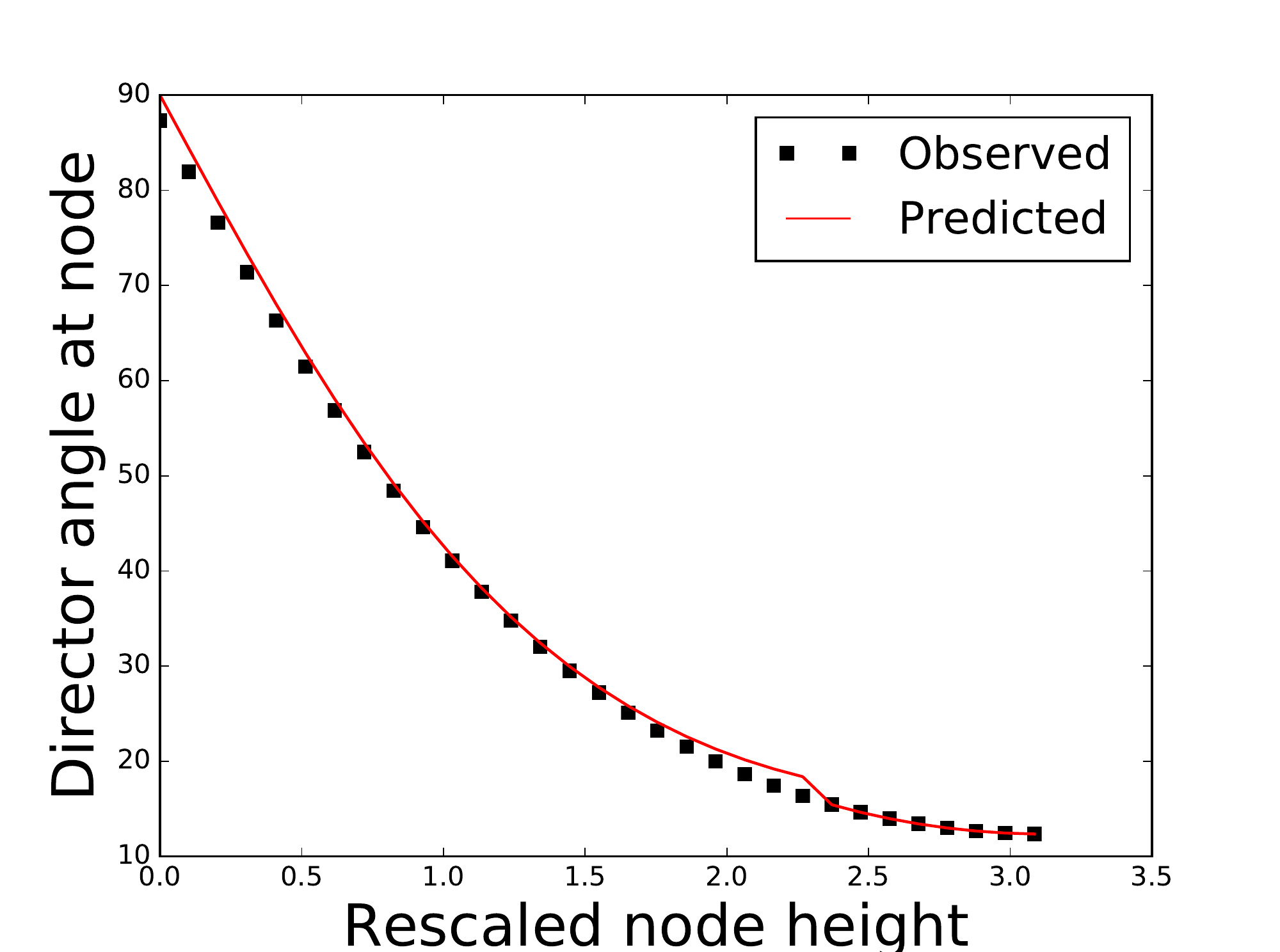}
}
{\includegraphics[height=4cm]{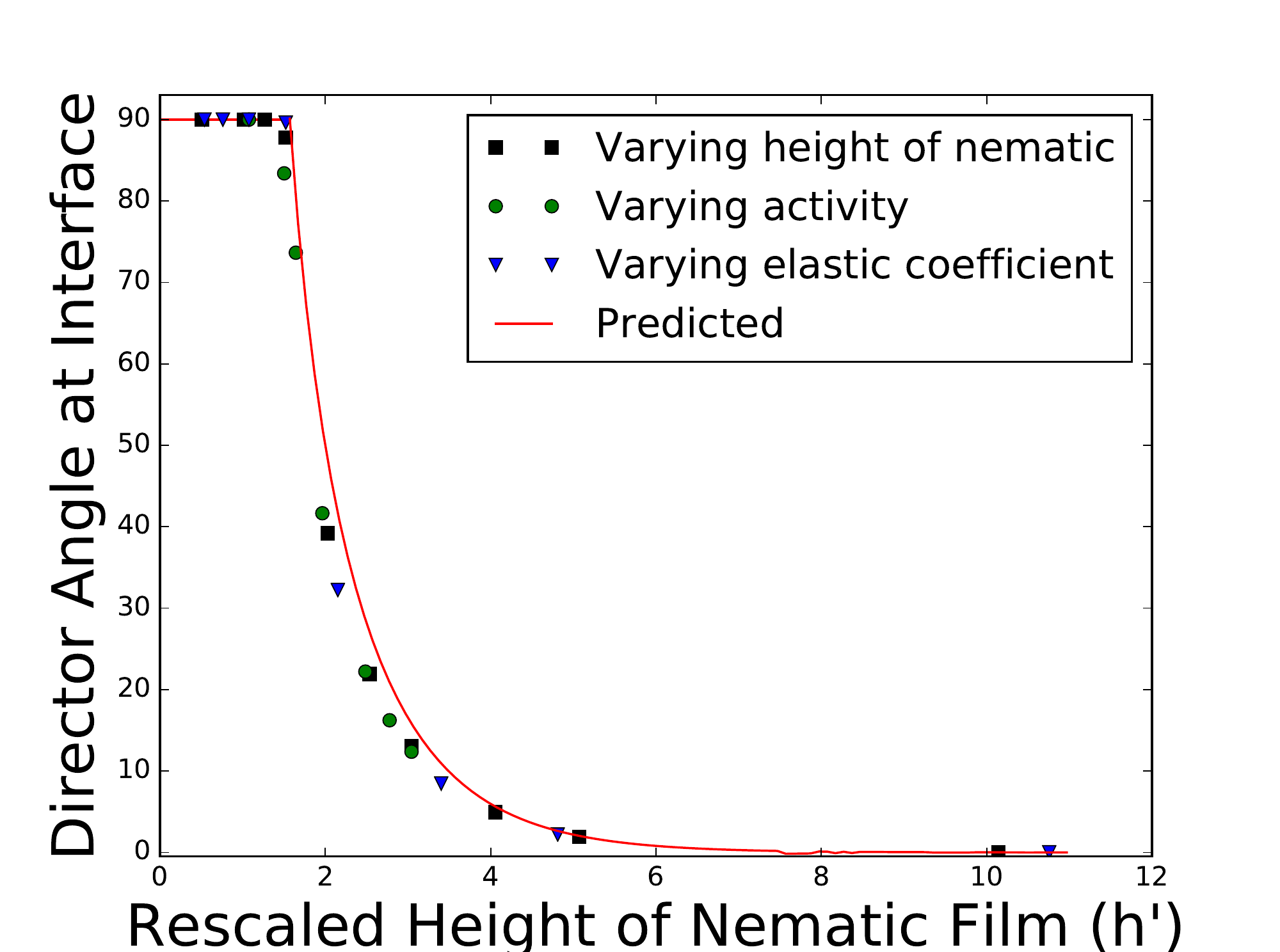}
}
{\includegraphics[height=4cm]{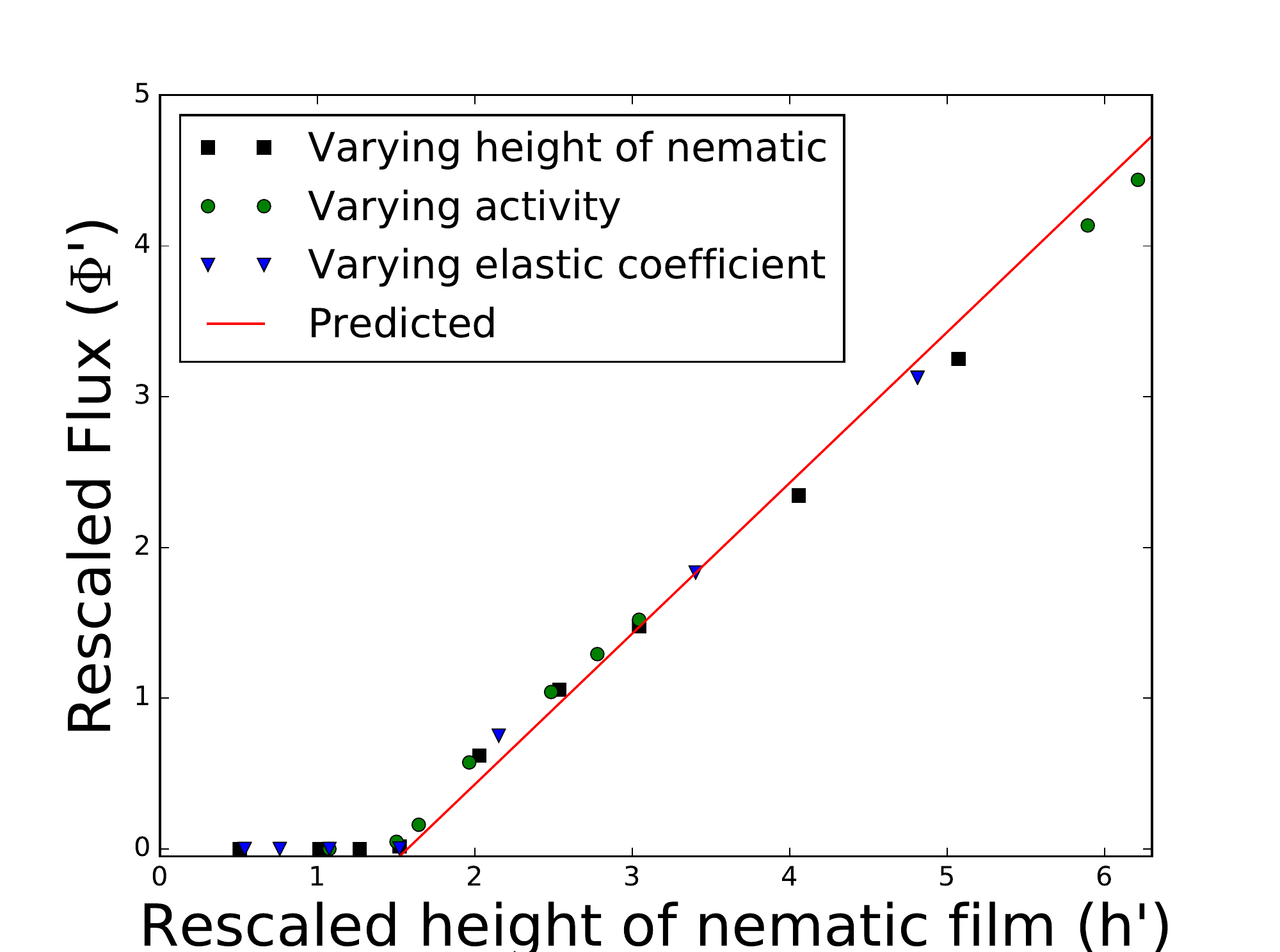}
}
\caption{(a) An example of the 1D geometry. The nematic phase is shown white and the isotropic phase blue. The black lines represent the scaled director $S\mathbf{n}$. (b) Director angle $\theta$ of an individual realisation as a function of $y'$ for the case $h'=3.04$. (c) Interface director angle $\thetaI$ and (d) rescaled volumetric flux $\Phi'$ as function of $h'$. The transition is clearly visible at $h'=\frac{\pi}{2}$.}
\label{fig:transition}

\end{center}
\end{figure*}

Nematic fluids are composed of head-tail-symmetric rodlike constituents that exhibit orientational ordering. We consider a 2D spatial domain occupying the $xy$-plane and assume that the director orientation is always within this plane, specified by $\mathbf{n}=(\cos\theta,\sin\theta)$. The appropriate order parameter is a symmetric, traceless tensor~\cite{DeGennesBook}:
\begin{equation}
\mathbf{Q}=S\begin{pmatrix}
\cos 2\theta & \sin 2\theta \\
\sin 2\theta & -\cos 2\theta
\end{pmatrix}, \label{eqn:Qtensor}
\end{equation}
Here, $S$ is the degree of nematic order. $S=0$ represents the isotropic phase, and we choose $S=1$ for the nematic phase.

The active nematic fluid coexists with an isotropic fluid, and the mass of each is constant. A conserved scalar parameter $\phi$ denotes the relative density of each fluid at a given point. The free energy of the system is
\begin{multline}
\mathcal{F}=\int\Big\{\tfrac{1}{2}A\phi^{2}\left(1-\phi\right)^{2} + \tfrac{1}{2}C\left(\phi^{2}-\tfrac{1}{2}Q_{\alpha\beta}Q_{\alpha\beta}\right)^{2} \\
+\tfrac{1}{2}K\vert\nabla\phi\vert^{2}+\tfrac{1}{2}L\partial_{\kappa}Q_{\alpha\beta}\partial_{\kappa}Q_{\alpha\beta}\Big\}d^{2}\mathbf{r}, \label{eqn:freeenergy}
\end{multline}
where $A$, $C$, $K$ and $L$ are positive constants. The first term in the integral is the bulk energy of the binary fluid~\cite{ChaikinBook,Orlandini1995}, which has two equilibria at $\phi=0,1$. The second term couples the nematic order $S$ to $\phi$, so as to favour isotropic order in regions where $\phi=0$, and nematic ordering in regions where $\phi=1$. 
The third and fourth terms, which penalise gradients in $\phi$ and $\mathbf{Q}$ respectively, both contribute to the surface tension~\cite{Orlandini1995,Sulaiman2006}, and the fourth term also provides the nematic elasticity in the bulk. 
Assuming that $C$ is sufficiently large that $S$ closely tracks $\phi$, the interface arising between the bulk phases has a characteristic width $\xiI\approx\sqrt{(2L+K)/A}$ and surface tension $\gamma\approx(2L+K)/(6\xiI)$.

The dynamical evolution of $\phi$, $\mathbf{Q}$, mass density $\rho$, and velocity $\mathbf{u}$ are governed by the equations~\cite{Cahn1958,BerisBook}
\begin{eqnarray}
\partial_{t}\phi+\partial_{\beta}(\phi u_{\beta})&=&M\nabla^{2}\mu, \label{eqn:cahnHilliard} \\
\left(\partial_{t}+u_{\kappa}\partial_{\kappa}\right)Q_{\alpha\beta}&=&Q_{\alpha\kappa}\epsilon_{\kappa\beta}\omega+\Gamma H_{\alpha\beta} \label{eqn:berisEds},\\
\partial_{t}\rho+\partial_{\beta}(\rho u_{\beta})&=&0,      \label{eqn:continuity} \\
\partial_{t}(\rho u_{\alpha})+\partial_{\beta}(\rho u_{\alpha}u_{\beta})&=&\partial_{\beta}\Big[\eta(\partial_{\beta}u_{\alpha}+\partial_{\alpha}u_{\beta}) -p\delta_{\alpha\beta}\nonumber\\
&&-\zeta Q_{\alpha\beta}\nonumber+Q_{\alpha\kappa}H_{\kappa\beta}-H_{\alpha\kappa}Q_{\kappa\beta}\Big]\nonumber\\
&&-H_{\kappa\lambda}\partial_{\alpha}Q_{\kappa\lambda}-\phi\partial_{\alpha}\mu.\label{eqn:navierStokes} 
\end{eqnarray}
where $M$ and $\Gamma$ are mobility constants, $\eta$ is the isotropic dynamic viscosity, and $p=\rho/3$ the isotropic pressure. $\omega=\partial_{x}u_{y}-\partial_{y}u_{x}$ is the vorticity, $\mu=\delta\mathcal{F}/\delta\phi$ is the chemical potential, and 
\begin{equation}
\begin{split}
H_{\alpha\beta}&=\frac{1}{2}\left(\frac{\delta\mathcal{F}}{\delta Q_{\kappa\kappa}}\delta_{\alpha\beta}-\frac{\delta\mathcal{F}}{\delta Q_{\alpha\beta}}-\frac{\delta\mathcal{F}}{\delta Q_{\beta\alpha}}\right) \\
&=L\nabla^{2}Q_{\alpha\beta}+C(\phi^{2}-\tfrac{1}{2}Q_{\kappa\lambda}Q_{\kappa\lambda})Q_{\alpha\beta} \label{eqn:molField}
\end{split}
\end{equation}
is the nematic molecular field. $\zeta$ is the strength of activity, corresponding to extensile activity when positive and contractile when negative~\cite{Simha2002,Davide2007}. 
We have omitted flow-aligning terms~\cite{BerisBook} in Eqns.~(\ref{eqn:berisEds},\ref{eqn:navierStokes}); we assume a pure flow-tumbling nematic to simplify the analytical calculations.

We use a hybrid simulation method~\cite{Davide2007,Tjhung2012} in which Eqns.~(\ref{eqn:cahnHilliard},\ref{eqn:berisEds}) are solved by finite differences, while the lattice Boltzmann method is applied to Eqns.~(\ref{eqn:continuity},\ref{eqn:navierStokes}). 
At the base of the simulation box, $y=0$, we apply the boundary conditions $u_{x}=0$ (no slip), and $\phi=S=1$ and $\theta=\thetaS$ (strong anchoring) to represent the substrate. 
At the top, we apply $\partial_{y}u_{x}=0$ and $\phi=S=0$, {\it i.e.} the system is taken to be open with a large body of isotropic fluid above the film. 
In the $x$ direction, we apply periodic boundary conditions. The system is initialised with nematic fluid $(\phi=S=1)$ in the region $y<h$, and isotropic fluid $(\phi=S=0)$ for $y>h$, $h$ thus being the thickness of the nematic film. In the nematic region, we initialise the director at a prescribed angle with random variations uniformly distributed over a given range ($\pm 9^{\circ}$ unless otherwise stated). Throughout this article we take $\Gamma=0.1$, $M=0.1$, $\rho=40$ (the fluid is near-incompressible), $\eta=26.67$ and $C=0.5$. Other parameters are varied, as described in the following sections (the values are listed in the relevant figure captions). In particular, for the 1D study of the no flux/flux transition (Figs.~\ref{fig:transition},\ref{fig:varySubstrate}) we vary height of nematic, activity, elastic constant and substrate anchoring angle. For the 2D simulations, we either vary surface tension and activity (phase diagram), or only the substrate anchoring angle.

\section{Approach and results}
\subsection{One dimension}
We first consider a 1D model in which $u_{y}$ and all $x$-derivatives are assumed to equal zero. This model is analytically solvable, and can be used to elucidate the basic principles by which active anchoring leads to film flow. To this end we keep the simulation box narrow in the $x$-direction, as depicted in Fig.~\ref{fig:transition}(a). We work with extensile activity, but the same principles will apply for contractile activity with the opposite type of substrate anchoring.

Assuming a steady state, Eqn.~(\ref{eqn:navierStokes})$_{x}$ integrates to
\begin{equation}
0 = \eta\partial_{y}u_{x} - \zeta Q_{xy} + 2(Q_{xx}H_{xy}-Q_{xy}H_{xx}), \label{eqn:NS1D}
\end{equation}
while taking $Q_{xx}$(\ref{eqn:berisEds})$_{xy}-Q_{xy}$(\ref{eqn:berisEds})$_{xx}$ gives
\begin{equation}
0 = -S^{2}\partial_{y}u_{x} + \Gamma(Q_{xx}H_{xy}-Q_{xy}H_{xx})\label{eqn:BE1D}.
\end{equation}
Using Eqns.~(\ref{eqn:Qtensor},\ref{eqn:molField}), we obtain $Q_{xx}H_{xy}-Q_{xy}H_{xx}=2LS\left(S\partial_{yy}\theta+2\partial_{y}\theta\partial_{y}S\right)$. Eliminating $\partial_{y}u_{x}$ from Eqns.~(\ref{eqn:NS1D},\ref{eqn:BE1D}) thus gives
\begin{equation}
S\partial_{y'y'}\theta + 2\partial_{y'}S\partial_{y'}\theta = \tfrac{1}{2}S\sin 2\theta, \label{eqn}
\end{equation}
where $y'=y/y_{0}$ is a non-dimensionalised length with $y_{0}=\sqrt{L(\Gamma\eta+2S^{2})/\zeta}$ as an active lengthscale. Typically, $y_{0}\gg\xiI$ (for typical values  $L=0.005$, $K=0.01$, A=$0.08$ and $\zeta=0.0001$, we have $\xiI=0.5$ and $y_{0}=15.3$) so we can assume that $\partial_{y'}S$ is very large at $h'=h/y_{0}$, and very small elsewhere. 
Thus we have the differential equation $\partial_{y'y'}\theta  = \tfrac{1}{2}\sin 2\theta$, subject to the boundary condition $\partial_{y'}\theta(h')=0$, which integrates to 
\begin{equation}
(\partial_{y'}\theta)^{2}=\cos^{2}\thetaI-\cos^{2}\theta,  \label{eqn:thetaInt}
\end{equation}
where $\thetaI=\theta(h')$ is the interfacial director angle. This equation has the same form as that describing an elastic column with a loaded end~\cite{LoveBook}, and the solution is
\begin{equation}
h'-y'=\K-\F\left[\arcsin\left(\frac{\cos\theta}{\cos\thetaI}\right)\right], \label{eqn:solution}
\end{equation}
where $\mathsf{K}$ and $\mathsf{F}$ denote complete and incomplete elliptic integrals of the first kind. $\thetaI$ is a free boundary value; to determine its relation to $h'$, we use the fixed boundary condition $\theta(0)=\thetaS$ in Eqn.~(\ref{eqn:solution}). For the case $\thetaS=\tfrac{\pi}{2}$ we obtain
\begin{equation}
h'=(2m+1)\mathsf{K}_{\cos\thetaI},\;\;\;m=0,1,2...  \label{eqn:heightRelation}
\end{equation}
and Eqn.~(\ref{eqn:solution}) simplifies to
\begin{equation}
\cos\theta=(-1)^{m}\cos\thetaI\sn(y'), \label{eqn:angleEquation}
\end{equation}
where $\textrm{sn}$ is the Jacobi elliptic sine. The corresponding solution has $m+1$ inflection points. Fig.~\ref{fig:transition}(b) compares Eqn.~(\ref{eqn:angleEquation}) ($m=0$) against simulation for $h'=3.04$.

Since $\mathsf{K}$ has a minimum value of $\tfrac{\pi}{2}$, occurring at zero elliptic modulus, 
a solution for a given $m$ is viable only for $h'>(m+\tfrac{1}{2})\pi$. However, even for very large $h'$, 
we never observe solutions with $m>0$ as the eventual steady state in the simulations. 
Even if the texture (director pattern) is initialised with multiple inflections, these will unwind to produce the $m=0$ state.
We therefore conclude that these higher solutions are unstable, as is the case for a loaded column~\cite{LoveBook}.

\begin{figure}[h!]
\begin{center}
{\includegraphics[height=4.8cm]{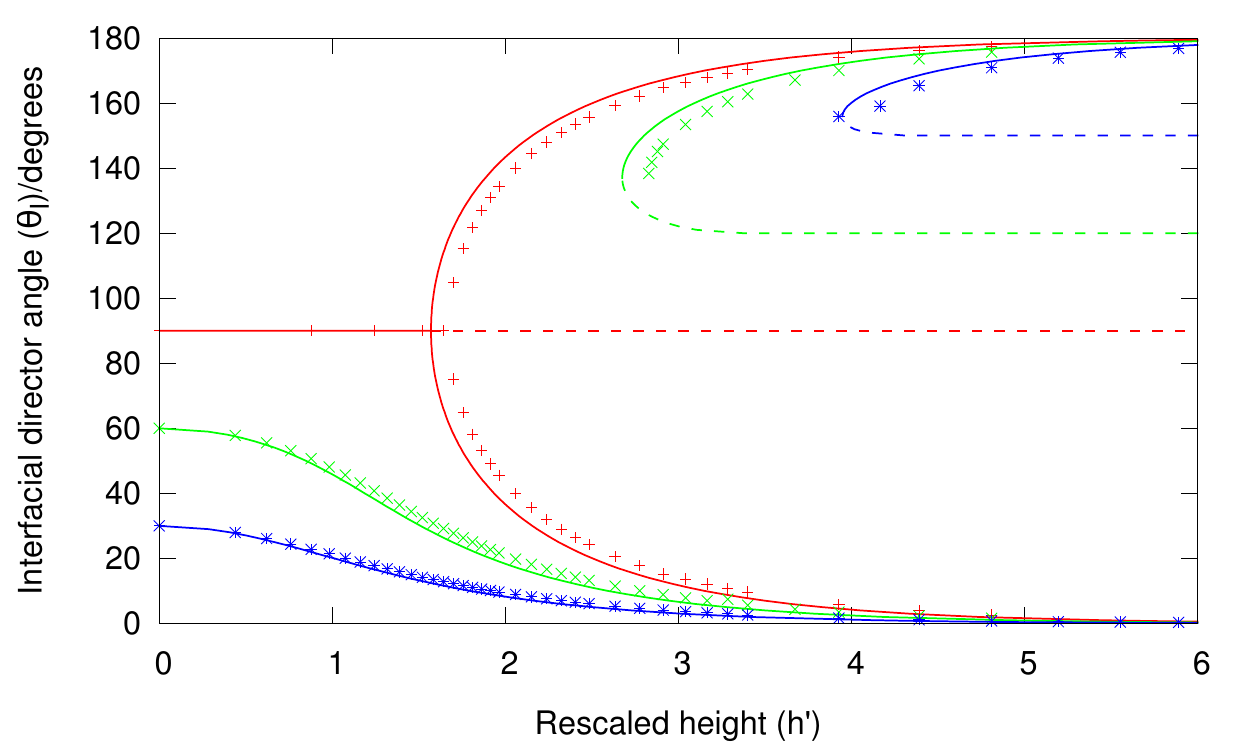}
}
{\includegraphics[height=4.8cm]{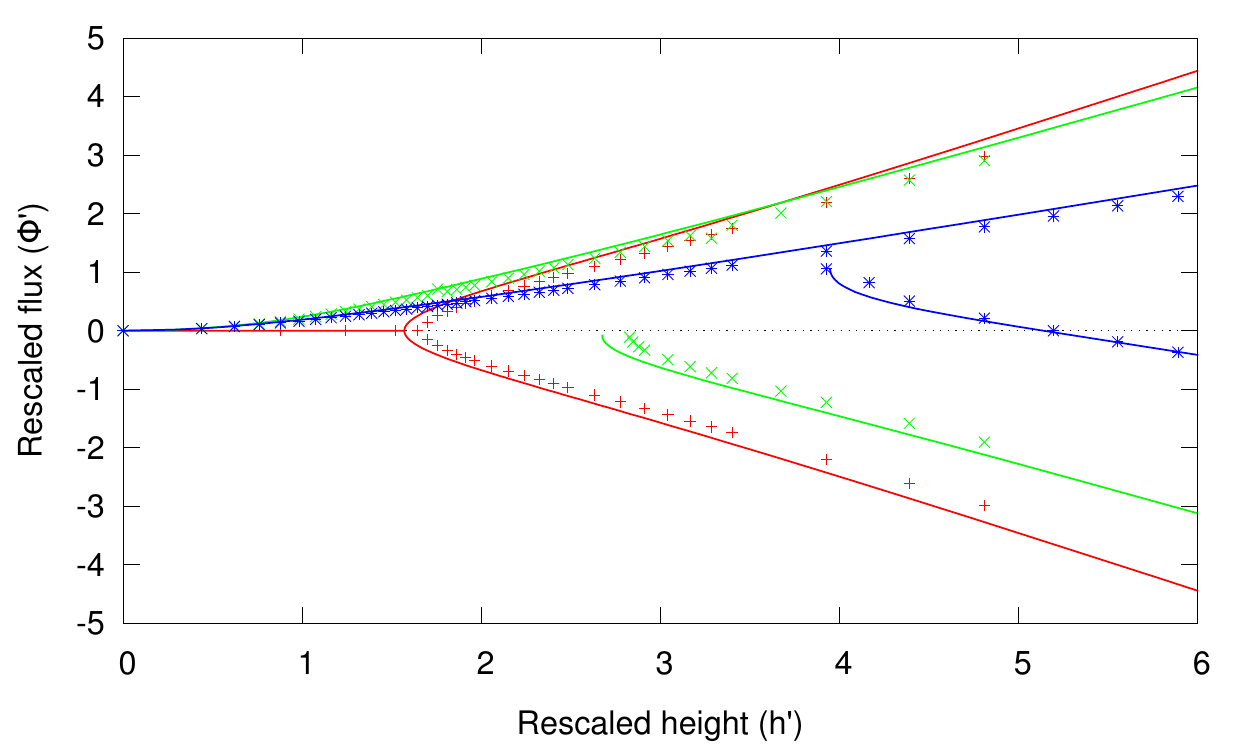}
}
{\includegraphics[height=4.8cm]{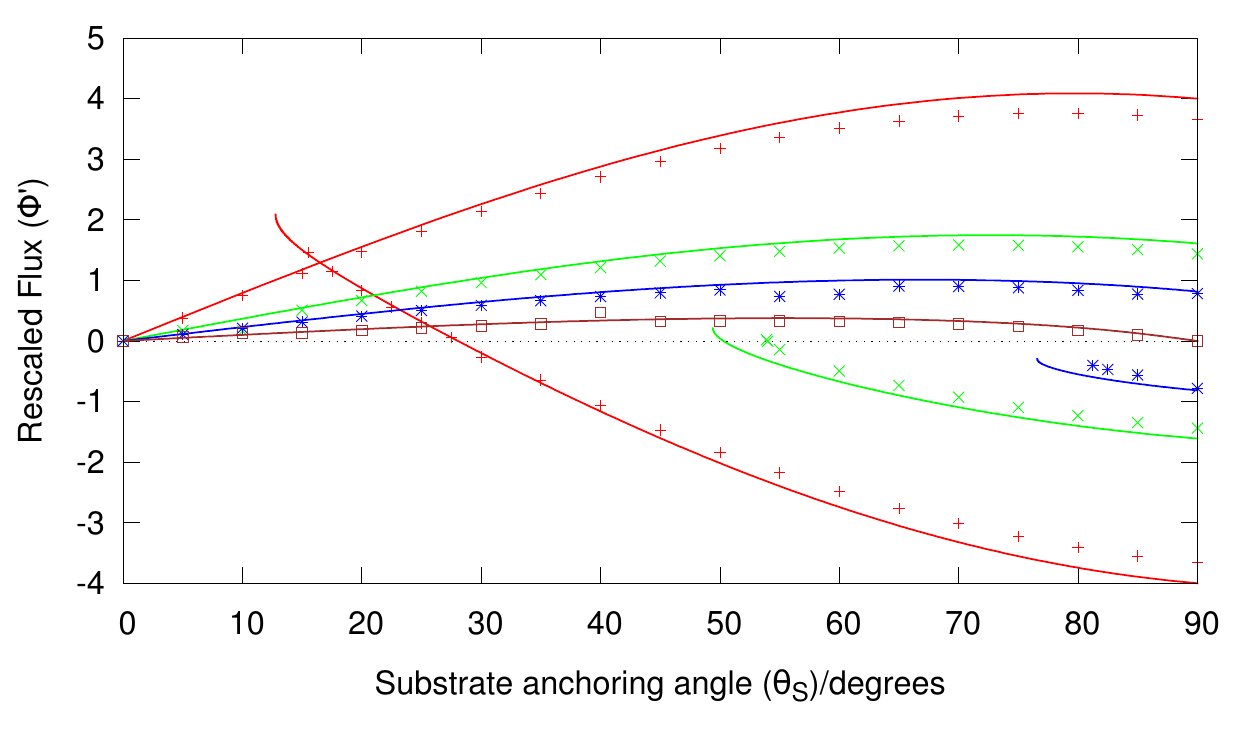}
}
\caption{(a) $\thetaI$ plotted against $h'$ for $\thetaS = 90^{\circ}$ (red `$+$'), $60^{\circ}$ (green `$\times$') and $30^{\circ}$ (blue `$*$'), for both the ``positive'' and ``negative'' solutions. The curves show the theoretical prediction (becoming dotted where the solution is unstable) and the markers show the simulation results. In all simulations $L=0.005$ and $h=30$, while $\zeta$ was varied to change $h'$. (b) $\Phi'$ for the same cases shown in (a). Only stable solutions are shown. (c) $\Phi'$ plotted against $\thetaS$ for $h'=5.56$ (red `$+$'s), $3.04$ (green `$\times$') $2.15$ (blue `$*$'), and $1.24$ (brown squares) (no ``negative'' solutions exist for this last case).}
\label{fig:varySubstrate}
\end{center}
\end{figure}

Analogously to the loaded column, we see an ``unbuckled'' state for $h'<\tfrac{\pi}{2}$, for which the director is vertical ($\theta=90^{\circ}$) throughout the film. 
At $h'=\tfrac{\pi}{2}$, there is a bifurcation into two possible ``buckled'' states (which are equivalent up to a reflection). 
The transition is similar to that described in~\cite{Voituriez2005}, for a film between two strongly-anchoring substrates, but occurs at half the film thickness, as one should expect for symmetry reasons. $\thetaI$ moves away from $90^{\circ}$ as $h'$ is further increased, and tends to $0$ (or equivalently $180^{\circ}$) for large $h'$, this being the planar alignment favoured by active anchoring.

Since $h'$ is scaled by the active lengthscale $l_{0}$, it depends not only on $h$, but also on $\zeta$ and $L$. Fig.~\ref{fig:transition}(c) 
shows the simulation results obtained by varying each of these three variables; good agreement is found with Eqn.~(\ref{eqn:heightRelation}) 
(only one bifurcation branch is shown). We note that a contractile active nematic on a planar-anchoring substrate would exhibit identical results but with $\theta\rightarrow\theta-\tfrac{\pi}{2}$ as can be shown based on Eqns.~(\ref{eqn:Qtensor}--\ref{eqn:navierStokes}).
Note however, that this correspondence is only exact for the pure flow-tumbling case adopted in this article. For a system with a non-zero flow-aligning parameter the situation is more complicated, as, for instance, extensile materials can spontaneously flow when they are sandwiched between two solid walls with tangential thermodynamic anchoring~\cite{Davide2007}.

Non-uniformity of nematic texture produces active forces, so the transition at $h'=\tfrac{\pi}{2}$ also demarcates non-flowing and flowing states. For a measure of flow, we use the volumetric flux across the film, defined 
as $\Phi=\int_{0}^{h}u_{x}dy$. We introduce the non-dimensionalised variables $\Phi'=\Phi/(2\Gamma L)$ and $u_{x}'=u_{x}y_{0}/(2\Gamma L)$. From Eqn.~(\ref{eqn:BE1D}) we find that $\partial_{y'}u_{x}'=\partial_{y'y'}\theta$. Thus, $u'=\partial_{y'}\theta(y')-\partial_{y'}\theta(0)$ and
\begin{equation}
\begin{split}
\Phi'=\thetaI-\frac{\pi}{2}+h'\cos\thetaI, \label{eqn:flux}
\end{split}
\end{equation}
where we have made use of Eqn.~(\ref{eqn:thetaInt}). Fig.~\ref{fig:transition}(d) compares this predicted $\Phi'$ against $h'$ with simulation results and shows
good agreement.
\begin{figure*}[htp]
\begin{center}
{\includegraphics[width=5.6cm]{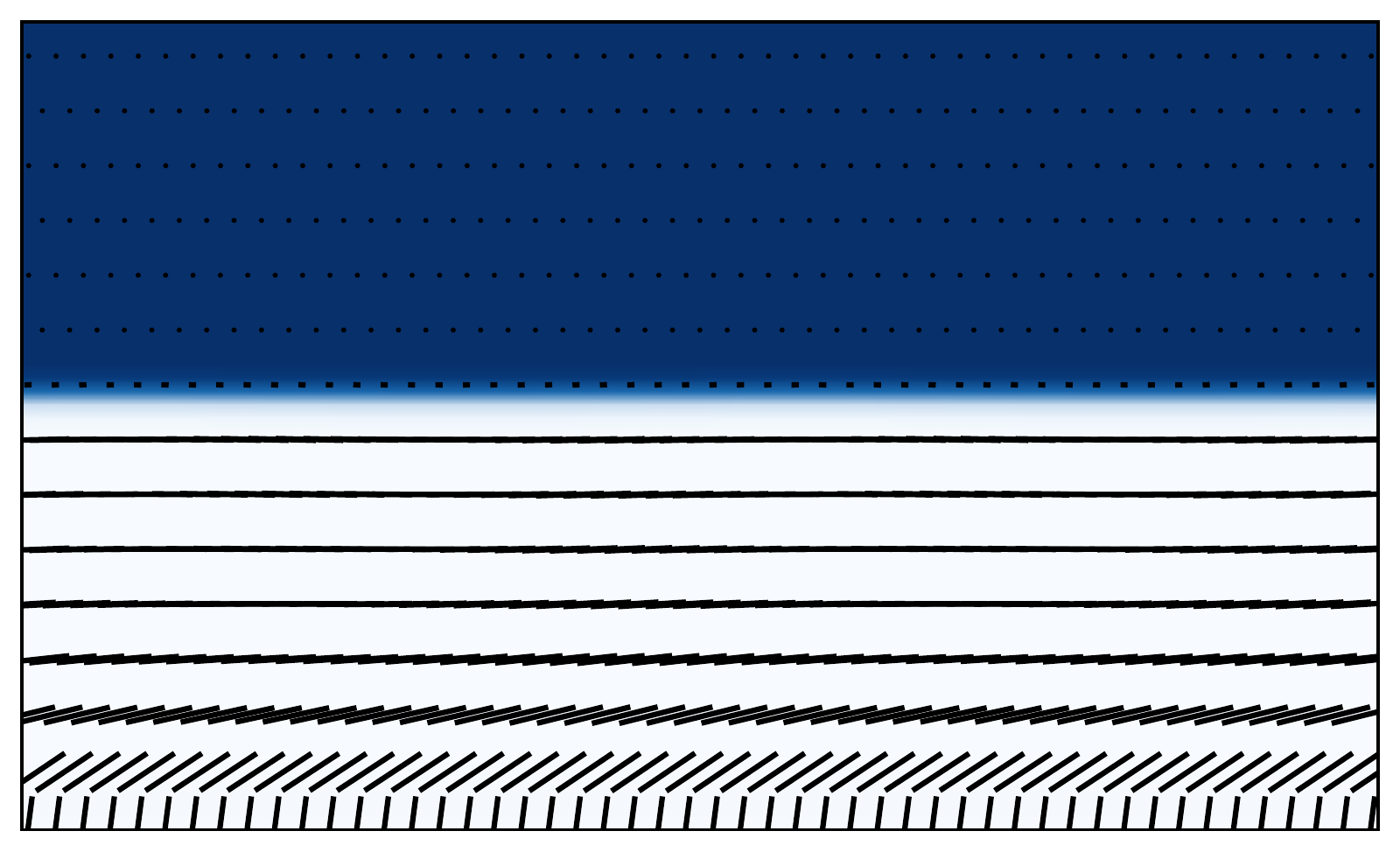}
}
\hfill
{\includegraphics[width=5.6cm]{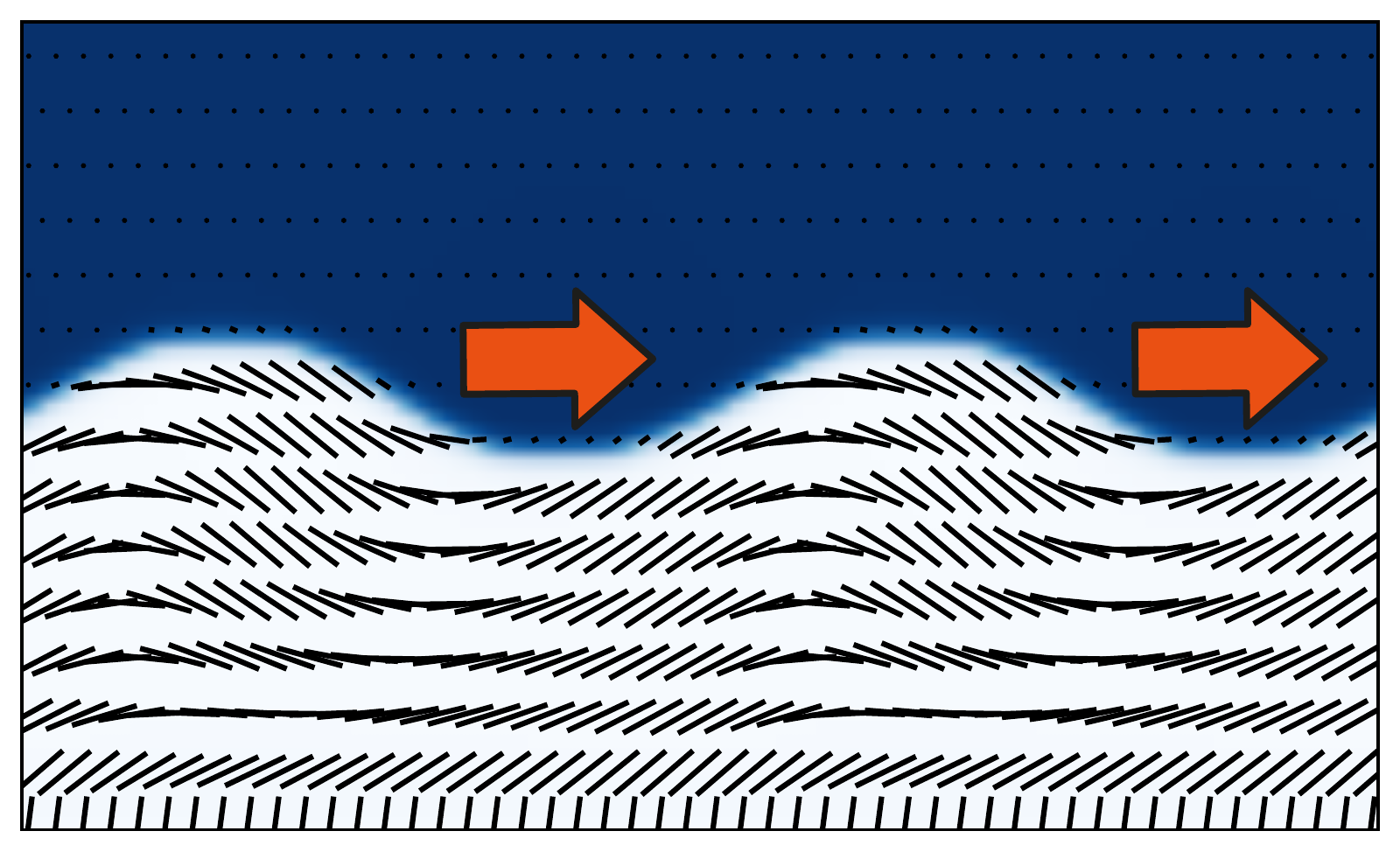}
}
\hfill
{\includegraphics[width=5.6cm]{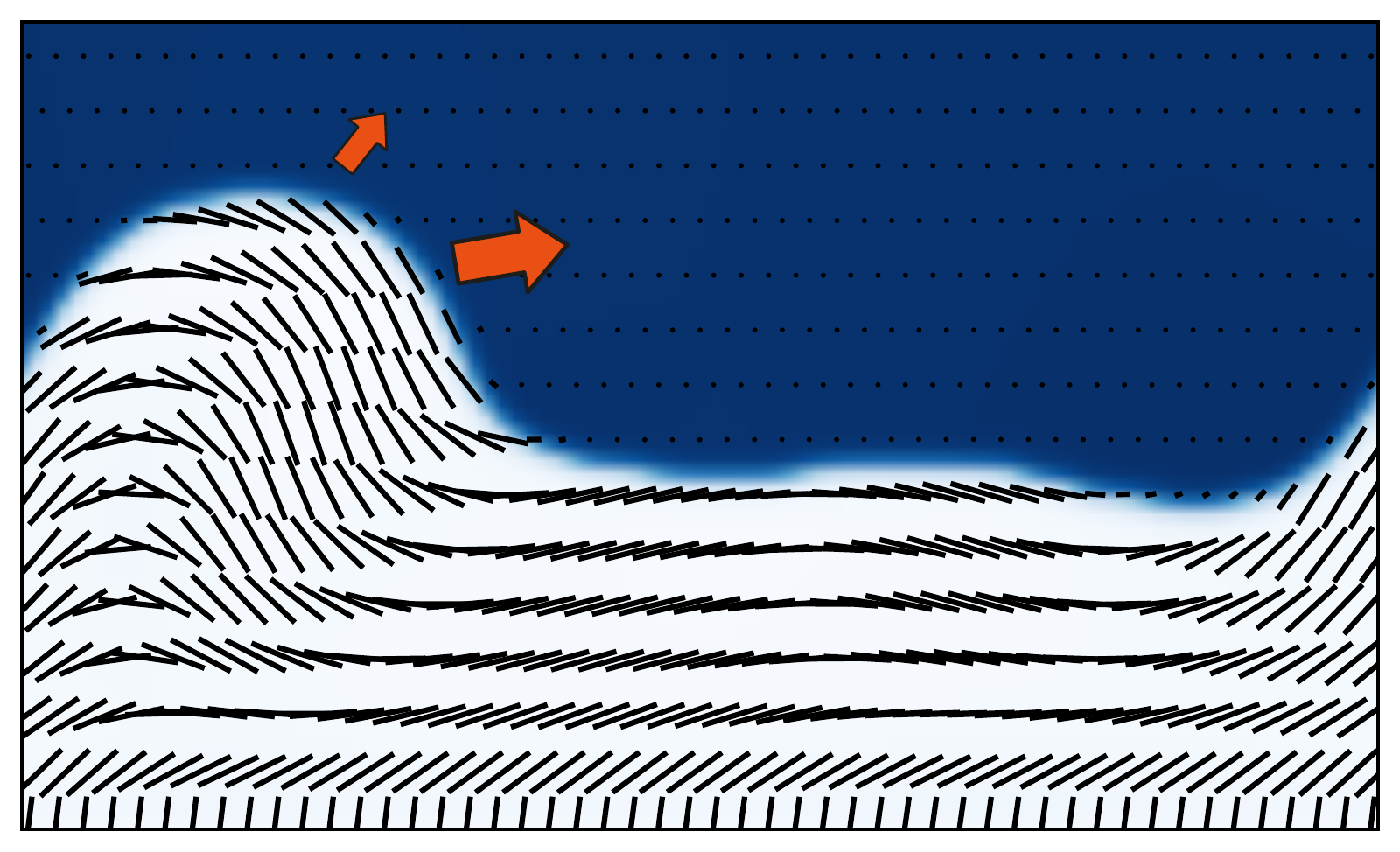}
}
\\
{\includegraphics[width=5.6cm]{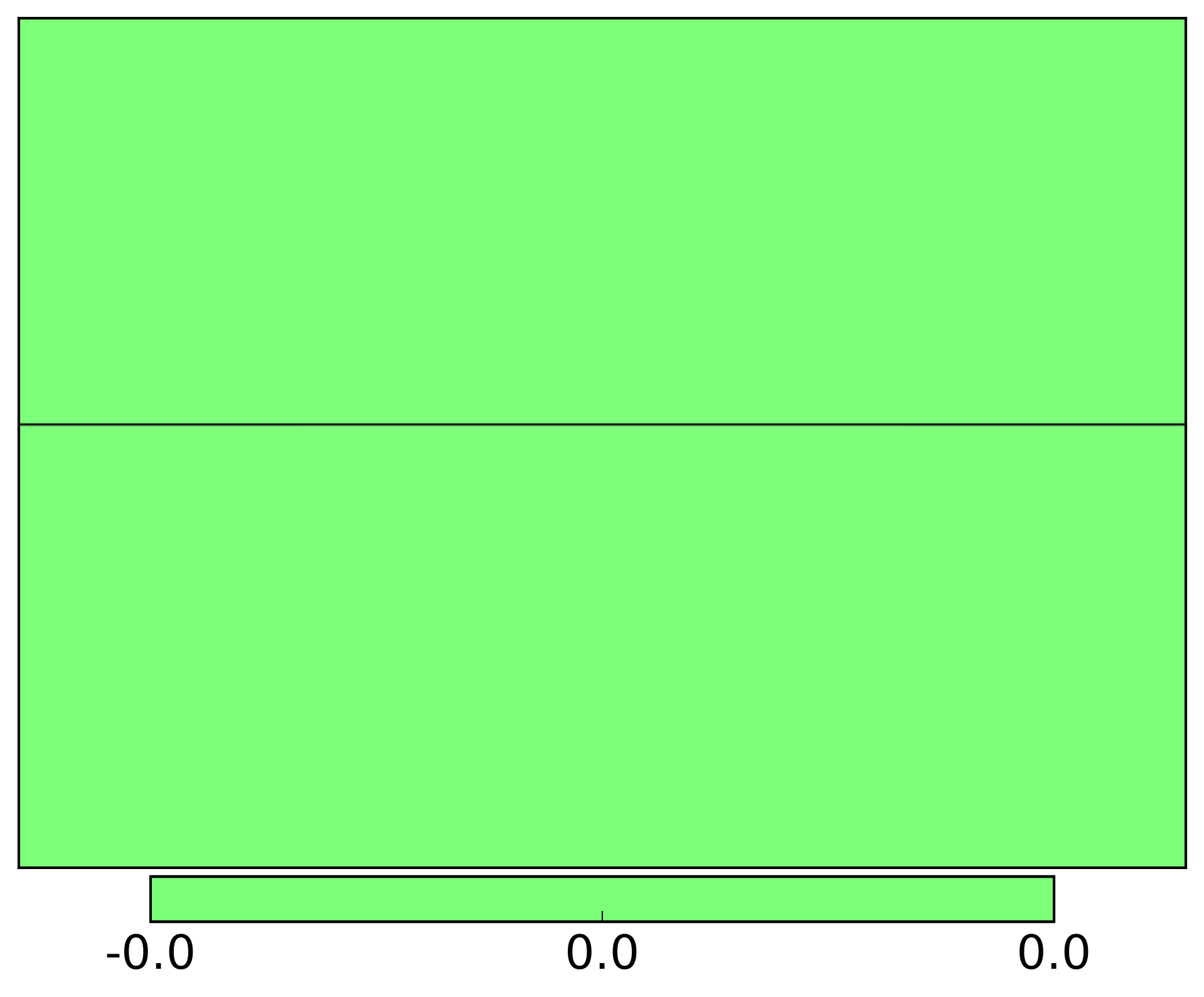}
}
\hfill
{\includegraphics[width=5.6cm]{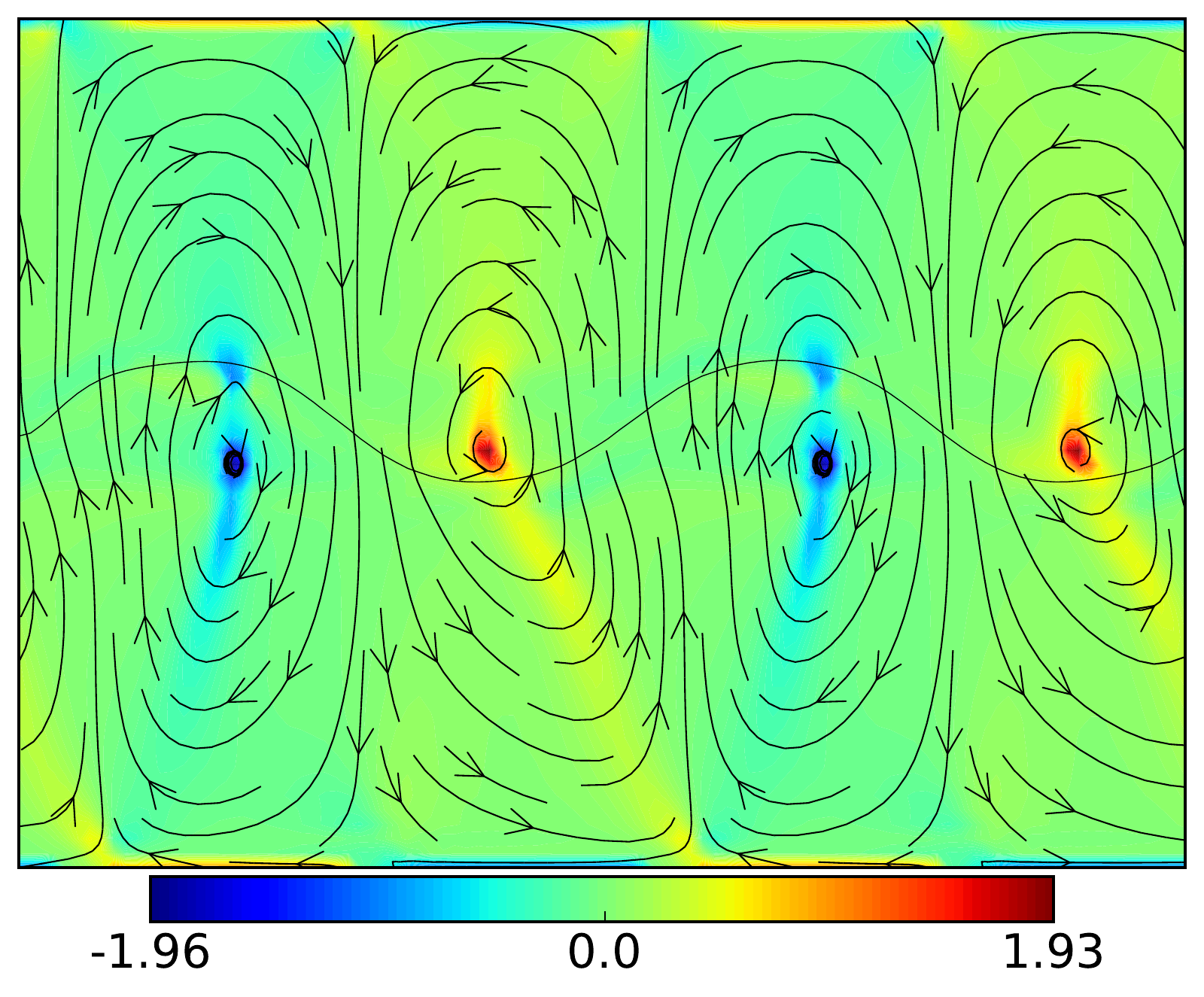}
}
\hfill
{\includegraphics[width=5.6cm]{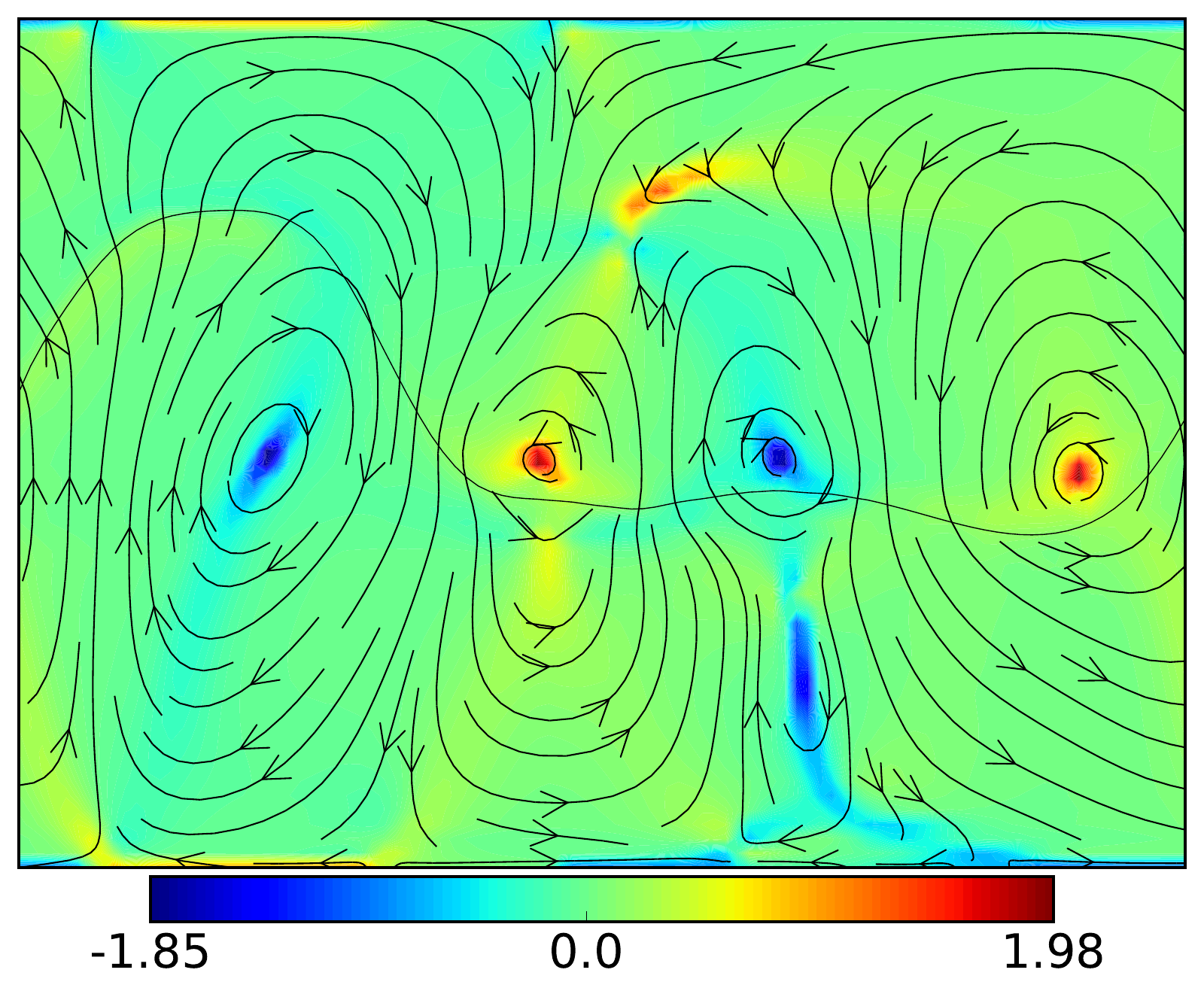}
}
\caption{Examples of possible 2D states. From left to right: flat interface, steady travelling wave, unstable interface. The upper plots show the nematic phase in white and the isotropic phase in blue. The black lines show the scaled nematic director $S\mathbf{n}$, and orange arrows indicate the motion of the interface. In the lower plots, the colouring corresponds to vorticity and the streamplot is of the deviatoric velocity field $\mathbf{u}(x,y)-\overline{u}_{x}(y)\hat{\mathbf{x}}$, where $\overline{u}_{x}(y)$ is the $x$-averaged horizontal velocity for a given value of $y$.
}
\label{fig:2Dexamples}
\end{center}
\end{figure*}

We now consider the case where the substrate anchoring angle $\thetaS$ differs from vertical. Using this boundary condition in Eqn.~(\ref{eqn:solution}), we derive the relation
\begin{equation}
h'=\K-\F\left[\arcsin\left(\frac{\cos\thetaS}{\cos\thetaI}\right)\right]
\end{equation}
(solutions with additional inflection points do exist, but again prove unstable). 
The non-vertical anchoring breaks reflectional symmetry, so there are two distinct solutions: a ``positive'' solution in 
which $\cos\thetaI$ and $\cos\thetaS$ have the same sign, and a ``negative'' solution in which the signs differ. 
Eqn.~(\ref{eqn:thetaInt}) shows that $\vert\cos\thetaI\vert\geq\vert\cos\thetaS\vert$ must hold in order for a real solution to exist. 
For the ``positive'' case, $\thetaI$ converges on $\thetaS$ when $h'=0$, and there is non-zero flow for all non-zero $h'$. 
For the ``negative'' case, $h'(\theta_I)$ has a minimum value, below which no solutions exist. Above this critical value, there are two possible solutions, but only the one with greater $\vert\cos\thetaI\vert$ is stable. Fig.~\ref{fig:varySubstrate}(a) shows $\thetaI$ against $h'$ for $\thetaS=30^{\circ}$ and $60^{\circ}$, compared with $\thetaS=90^{\circ}$.

In a generalisation of Eqn.~(\ref{eqn:flux}), the volumetric flux is given by 
\begin{equation}
\begin{split}
\Phi'=\thetaI-\thetaS+h'\cos\thetaI\sqrt{1-\frac{\cos^{2}\thetaS}{\cos^{2}\thetaI}}. \label{eqn:fluxInclined}
\end{split}
\end{equation}
Fig.~{\ref{fig:varySubstrate}}(b) shows $\Phi'$ against $h'$, again for $\thetaS=30^{\circ},60^{\circ},90^{\circ}$. The flow for the ``negative'' state is usually opposite in direction to that of the ``positive'', but for lower $\thetaS$ can flow in the same direction, as is shown for $\thetaS=30^{\circ}$. Fig.~{\ref{fig:varySubstrate}}(c) shows how $\Phi'$ varies with $\thetaS$ for constant values of $h'$. The maximum $\Phi'$ occurs for $\thetaS$ slightly below $90^{\circ}$ when $h'$ is high, moving to lower $\thetaS$ as $h'$ is reduced. Interestingly, the curve for highest $h'$ value ($h'=5.555$) shows the $\Phi'$ of the ``negative'' state not only becoming positive, but actually surpassing the $\Phi'$ of the ``positive'' state. However, in simulations of such cases, the inital director configuration must be chosen with care in order for the system to adopt the ``negative'' state. We thus conclude that such situations are unlikely to occur naturally.

\begin{figure}[h!]
\begin{center}
{\includegraphics[width=8cm]{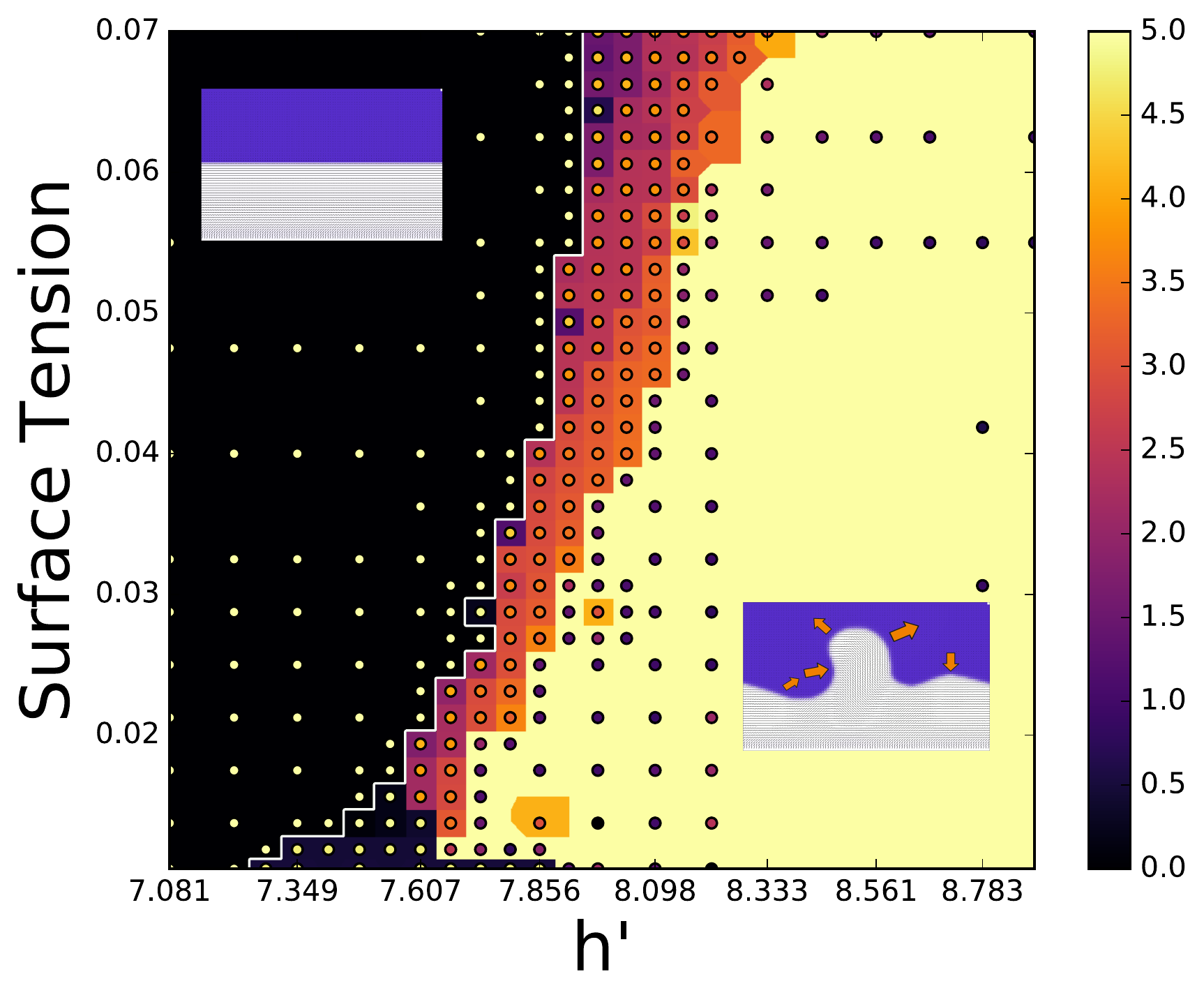}
}
\caption{Phase diagram showing the measured roughness (in lattice cells) of the interface with given activity and surface tension. The contour line pinpoints the transition from zero to non-zero roughness states. A colour corresponding to the top of the scale does not necessarily signify a roughness of $5.0$, but rather where the roughness diverges indefinitely or varies aperiodically. The insets show qualitative examples of states from the respective regions of the diagram. The $x$ axis gives values of $h'$, to provide comparison with the flow transition (which occurs at $h'=\tfrac{\pi}{2}$, below the range shown here), but note that only activity has been varied.}
\label{fig:phaseDiagram}
\end{center}
\end{figure}
\subsection{Two dimensions}
The 1D analysis assumes that the film remains translationally invariant along the $x$ axis. However, active nematics are orientationally unstable on sufficiently large scales~\cite{Simha2002,Ramaswamy2010}.
Specifically for an active film, but with different boundary conditions, translational instabilities have been demonstrated~\cite{Sankararaman2009}. To investigate the effects of such instabilities in an active nematic exposed to both a solid substrate and a free interface, we now perform 2D simulations in which the simulation box spans many lattice sites along $x$. We find that, as a consequence of this bulk instability, there are parameters for which variation along $x$ occurs in both the director profile and in the height of the interface. These interfacial undulations mean that the surface tension, largely irrelevant in the 1D model, plays an important role in determining the behaviour.

Throughout the following we set $\thetaS=90^{\circ}$ and use a simulation box of at least $100\times 60$ lattice nodes. To vary $h'$ we change $\zeta$, while keeping $L=0.005$ and $h=30$. $h'$ is always above the no-flow/flow transition identified for 1D. We also vary $A$ and $K$ in such a way as to vary the surface tension while maintaining a constant interfacial width.

We can separate the behaviour of the film into three broad regimes, examples of which are depicted in Fig.~\ref{fig:2Dexamples}. Firstly, for low activity the interface remains flat (Fig.~\ref{fig:2Dexamples}(a)), though there can be small distortions in the orientational order within the active nematic slab.

In a second regime, occurring at intermediate activity, the interface exhibits a non-zero but constant roughness, and settles to a wave of finite amplitude in steady state (Fig.\ref{fig:2Dexamples}(b)). Here, the nematic texture and the interfacial wave do not change shape over time, but only translate steadily. In general, the interfacial wave does not travel at the same velocity as the fluid; as the surface tension or the amplitude of waves increases, the waves typically become slower relative to the fluid flow.

Finally, in a third regime occuring at high activity, the amplitude of interfacial oscillations fluctuates, either periodically or aperiodically (Fig.~\ref{fig:2Dexamples}(c)). If surface tension is low enough, the nematic film can even eject droplets or break up (SM, video 3).

Fig.~(\ref{fig:phaseDiagram}) provides a slice through the phase diagram based on many individual simulations with noisy initial conditions. The colouring represents the time-averaged amplitude of the interfacial roughness. The diagram shows the three regimes (flat interface, black; steady interfacial waves, purple to orange; irregular undulations, cream) corresponding to the representative examples in Fig.~\ref{fig:2Dexamples}, which we discussed above.

Starting from the flat-interface regime on the left of Fig.~(\ref{fig:phaseDiagram}), as activity is increased the film undergoes a continuous transition to non-zero roughness, indicating the onset of the steady surface waves. This transition line occurs at an $h'$ approximately $5$ times that at which the stationary to moving transition occurs ($\pi/2$). This $h'$ increases slightly as the surface tension is increased, in line with the intuition that augmenting the surface tension greatens the free energy cost associated with interface deformations, hence stabilising a flat interface. The steady wave regime occupies a narrow band close to the transition line, beyond which both the flow rate and interfacial roughness vary irregularly over time.

The timescale characterising the emergence of both non-zero roughness and aperiodic behaviour differs across the phase diagram. As usual in pattern formation problems, simulations closer to the transition line take longer to become unstable compared to those far in the unsteady regime.

The simulations shed light on how the interfacial instability arising in the present system is linked to the well known orientional bulk instability (the so-called generic instability)~\cite{Simha2002,Ramaswamy2010}. The latter instability leads to deformations in the orientational order, which in turn generate vorticity in the flow field (convection rolls, see Fig.~\ref{fig:2Dexamples}). These rolls pull and push at alternating positions at the interface, and if dominant over the restoring surface tension, lead to interfacial modulations. The modulated interface, in turn, displaces the centres of the convection rolls until force balance is achieved and both interface and convection rolls remain stationary (Fig.~\ref{fig:2Dexamples}b). However, our simulations show other possible scenarios. At high activity no balance is possible between active fluid motion and restoring interfacial forces (Fig.~\ref{fig:2Dexamples}c), and thus we observe irregular interfacial oscillations.

\begin{figure}[h!]
\begin{center}
{\includegraphics[width=4.2cm]{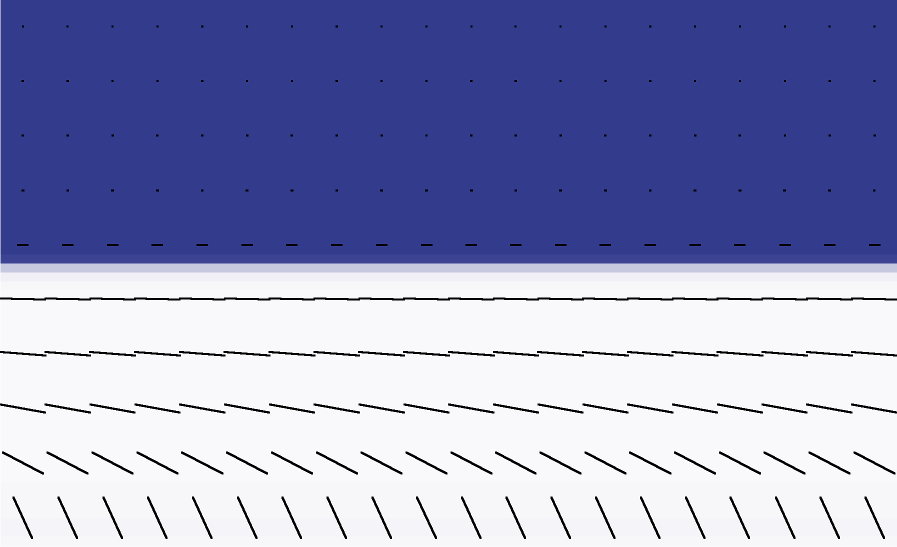}
}
\hfill
{\includegraphics[width=4.2cm]{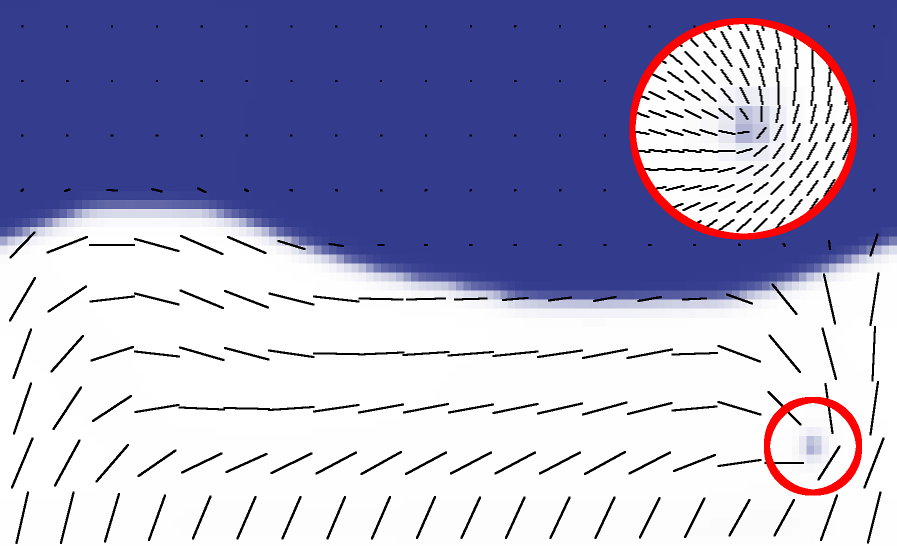}
}
\\
{\includegraphics[width=4.2cm]{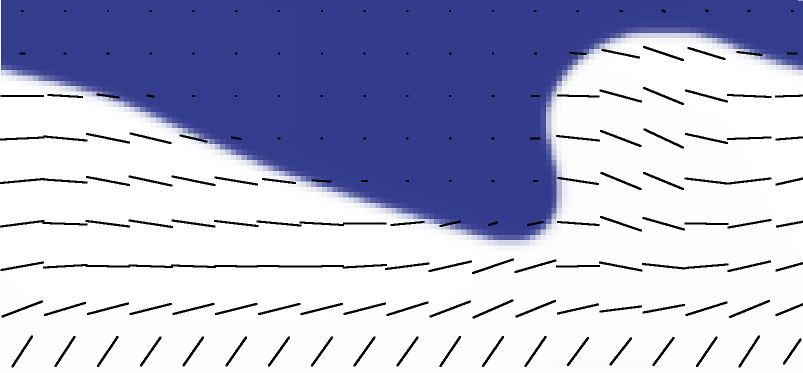}
}
\hfill
{\includegraphics[width=4.2cm]{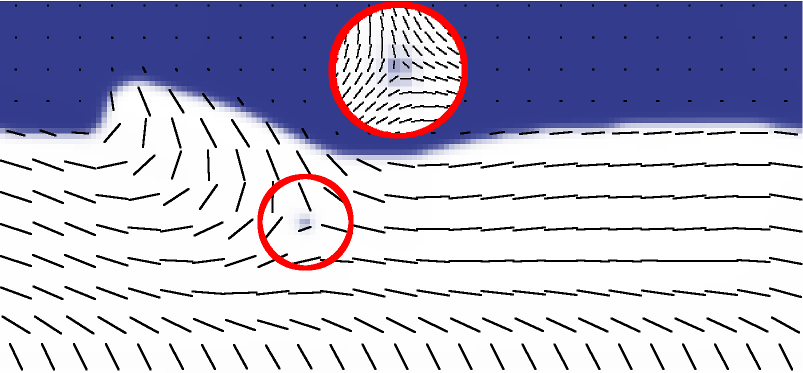}
}
\caption{Two examples of qualitatively different steady states arising for the same set of parameters. For the top row $h'=5.33$, while for the bottom row $h'=7.86$. The left-hand states are defect-free while in the states on the right a singular topological defect is trapped in the texture (topological charge $+1/2$ above and $-1/2$ below), leaving the interface with a diffuse topological charge of the opposite sign. The detail of the defects is expanded in the red circles.}
\label{fig:defects}
\end{center}
\end{figure}

The results presented so far were obtained starting from small fluctuations ($\pm 9^{\circ}$) from a uniform director configuration. Increasing the amplitude of these fluctuations allows the formation of nematic defects within the bulk of the nematic. The majority of these defects annihilate during the subsequent evolution, but in some cases a single nematic defect remains trapped in the nematic phase, even in the steady state. It has been shown that seemingly unpaired defects can exist in the bulk, provided that the interface has a diffuse topological charge of the opposite sign~\cite{Blow2014}. The presence of this topological charge can profoundly alter the morphology of the interface and in particular its curvature, and hence the steady states with trapped defects are qualitatively different to those states for the same system parameters but without trapped defects. Examples are shown in Figure~\ref{fig:defects}; the snapshots on the left are examples of defect-free states for two parameter sets, while those on the right show states with defects, for the same two parameter sets.

\section{Discussion}
\label{sec:discussion}
In this article we have demonstrated how the combination of thermodynamic and ``active'' anchoring (respectively determined by free energy and active flows) can drive persistent flow in an active nematic film. 
In one dimension our system is analogous to a loaded elastic rod, and in the case of extensile activity with homeotropic substrate anchoring (or likewise contractile activity with planar substrate anchoring), exhibits a ``buckling'' transition from a stationary to a spontaneously flowing state, whose direction depends on the initial conditions.
Conversely, in 2D, the generic bulk instability of active nematics~\cite{Simha2002} triggers interfacial undulations, and we have shown that these may result in either regular propagating surface waves, or in irregular patterns. 

Our work has potential implications for the understanding of biological flows and for the design of novel active microfluidic geometries. For example, Fig.~\ref{fig:varySubstrate}(c) shows how anchoring angle may be used to tailor flow properties of active films. Having $\thetaS$ slightly below $90^{\circ}$ is likely to be advantageous compared to $90^{\circ}$ exactly, firstly because the flow throughput $\Phi'$ is slightly higher, and secondly because the broken symmetry improves consistency of flow direction. Slanting the $\thetaS$ further away from $90^{\circ}$ will further improve this consistency, since the ``negative'' solutions become unviable, but at the expense of a diminishing throughput.

The existence of states with steady interfacial undulations also has potential applications, as such interfaces might be used for curvature-guided self-assembly~\cite{Grzybowski2001} and switchable diffraction gratings~\cite{Brown2009}.

\section*{Acknowledgements}
We thank Alexander Morozov and Julia Yeomans for helpful discussions and comments. The work was funded in part by EPSRC EP/J007404/1. M.A. acknowledges support from the Higgs Centre Visiting Researcher Programme. B.L. acknowledges funding by a Marie Sk{\l}odowska Curie Intra European Fellowship (G.A. no 654908) within Horizon 2020.

%\clearpage

\bibliography{refe}

\providecommand{\noopsort}[1]{}\providecommand{\singleletter}[1]{#1}%
\begin{thebibliography}{32}%
\makeatletter
\providecommand \@ifxundefined [1]{%
 \@ifx{#1\undefined}
}%
\providecommand \@ifnum [1]{%
 \ifnum #1\expandafter \@firstoftwo
 \else \expandafter \@secondoftwo
 \fi
}%
\providecommand \@ifx [1]{%
 \ifx #1\expandafter \@firstoftwo
 \else \expandafter \@secondoftwo
 \fi
}%
\providecommand \natexlab [1]{#1}%
\providecommand \enquote  [1]{``#1''}%
\providecommand \bibnamefont  [1]{#1}%
\providecommand \bibfnamefont [1]{#1}%
\providecommand \citenamefont [1]{#1}%
\providecommand \href@noop [0]{\@secondoftwo}%
\providecommand \href [0]{\begingroup \@sanitize@url \@href}%
\providecommand \@href[1]{\@@startlink{#1}\@@href}%
\providecommand \@@href[1]{\endgroup#1\@@endlink}%
\providecommand \@sanitize@url [0]{\catcode `\\12\catcode `\$12\catcode
  `\&12\catcode `\#12\catcode `\^12\catcode `\_12\catcode `\%12\relax}%
\providecommand \@@startlink[1]{}%
\providecommand \@@endlink[0]{}%
\providecommand \url  [0]{\begingroup\@sanitize@url \@url }%
\providecommand \@url [1]{\endgroup\@href {#1}{\urlprefix }}%
\providecommand \urlprefix  [0]{URL }%
\providecommand \Eprint [0]{\href }%
\providecommand \doibase [0]{http://dx.doi.org/}%
\providecommand \selectlanguage [0]{\@gobble}%
\providecommand \bibinfo  [0]{\@secondoftwo}%
\providecommand \bibfield  [0]{\@secondoftwo}%
\providecommand \translation [1]{[#1]}%
\providecommand \BibitemOpen [0]{}%
\providecommand \bibitemStop [0]{}%
\providecommand \bibitemNoStop [0]{.\EOS\space}%
\providecommand \EOS [0]{\spacefactor3000\relax}%
\providecommand \BibitemShut  [1]{\csname bibitem#1\endcsname}%
\let\auto@bib@innerbib\@empty
%</preamble>
\bibitem [{\citenamefont {Sanchez}\ \emph {et~al.}(2012)\citenamefont
  {Sanchez}, \citenamefont {Chen}, \citenamefont {DeCamp}, \citenamefont
  {Heymann},\ and\ \citenamefont {Dogic}}]{Sanchez2012}%
  \BibitemOpen
  \bibfield  {author} {\bibinfo {author} {\bibfnamefont {T.}~\bibnamefont
  {Sanchez}}, \bibinfo {author} {\bibfnamefont {D.~T.~C.}\ \bibnamefont
  {Chen}}, \bibinfo {author} {\bibfnamefont {S.~J.}\ \bibnamefont {DeCamp}},
  \bibinfo {author} {\bibfnamefont {M.}~\bibnamefont {Heymann}}, \ and\
  \bibinfo {author} {\bibfnamefont {Z.}~\bibnamefont {Dogic}},\ }\href
  {\doibase 10.1038/nature11591} {\bibfield  {journal} {\bibinfo  {journal}
  {Nature}\ }\textbf {\bibinfo {volume} {491}},\ \bibinfo {pages} {431}
  (\bibinfo {year} {2012})}\BibitemShut {NoStop}%
\bibitem [{\citenamefont {Prost}\ \emph {et~al.}(2015)\citenamefont {Prost},
  \citenamefont {J\"{u}licher},\ and\ \citenamefont {Joanny}}]{Prost2015}%
  \BibitemOpen
  \bibfield  {author} {\bibinfo {author} {\bibfnamefont {J.}~\bibnamefont
  {Prost}}, \bibinfo {author} {\bibfnamefont {F.}~\bibnamefont {J\"{u}licher}},
  \ and\ \bibinfo {author} {\bibfnamefont {J.-F.}\ \bibnamefont {Joanny}},\
  }\href {\doibase 10.1038/nphys3224} {\bibfield  {journal} {\bibinfo
  {journal} {Nat. Phys.}\ }\textbf {\bibinfo {volume} {11}},\ \bibinfo {pages}
  {111} (\bibinfo {year} {2015})}\BibitemShut {NoStop}%
\bibitem [{\citenamefont {Dunkel}\ \emph {et~al.}(2013)\citenamefont {Dunkel},
  \citenamefont {Heidenreich}, \citenamefont {Drescher}, \citenamefont
  {Wensink}, \citenamefont {B\"ar},\ and\ \citenamefont
  {Goldstein}}]{Dunkel2013}%
  \BibitemOpen
  \bibfield  {author} {\bibinfo {author} {\bibfnamefont {J.}~\bibnamefont
  {Dunkel}}, \bibinfo {author} {\bibfnamefont {S.}~\bibnamefont {Heidenreich}},
  \bibinfo {author} {\bibfnamefont {K.}~\bibnamefont {Drescher}}, \bibinfo
  {author} {\bibfnamefont {H.~H.}\ \bibnamefont {Wensink}}, \bibinfo {author}
  {\bibfnamefont {M.}~\bibnamefont {B\"ar}}, \ and\ \bibinfo {author}
  {\bibfnamefont {R.~E.}\ \bibnamefont {Goldstein}},\ }\href {\doibase
  10.1103/PhysRevLett.110.228102} {\bibfield  {journal} {\bibinfo  {journal}
  {Phys. Rev. Lett.}\ }\textbf {\bibinfo {volume} {110}},\ \bibinfo {pages}
  {228102} (\bibinfo {year} {2013})}\BibitemShut {NoStop}%
\bibitem [{\citenamefont {Golestanian}\ \emph {et~al.}(2005)\citenamefont
  {Golestanian}, \citenamefont {Liverpool},\ and\ \citenamefont
  {Ajdari}}]{Golestanian2005}%
  \BibitemOpen
  \bibfield  {author} {\bibinfo {author} {\bibfnamefont {R.}~\bibnamefont
  {Golestanian}}, \bibinfo {author} {\bibfnamefont {T.~B.}\ \bibnamefont
  {Liverpool}}, \ and\ \bibinfo {author} {\bibfnamefont {A.}~\bibnamefont
  {Ajdari}},\ }\href {\doibase 10.1103/PhysRevLett.94.220801} {\bibfield
  {journal} {\bibinfo  {journal} {Phys. Rev. Lett.}\ }\textbf {\bibinfo
  {volume} {94}},\ \bibinfo {pages} {220801} (\bibinfo {year}
  {2005})}\BibitemShut {NoStop}%
\bibitem [{\citenamefont {Giomi}\ \emph {et~al.}(2012)\citenamefont {Giomi},
  \citenamefont {Mahadevan}, \citenamefont {Chakraborty},\ and\ \citenamefont
  {Hagan}}]{Giomi2012}%
  \BibitemOpen
  \bibfield  {author} {\bibinfo {author} {\bibfnamefont {L.}~\bibnamefont
  {Giomi}}, \bibinfo {author} {\bibfnamefont {L.}~\bibnamefont {Mahadevan}},
  \bibinfo {author} {\bibfnamefont {B.}~\bibnamefont {Chakraborty}}, \ and\
  \bibinfo {author} {\bibfnamefont {M.~F.}\ \bibnamefont {Hagan}},\ }\href
  {\doibase 10.1088/0951-7715/25/8/2245} {\bibfield  {journal} {\bibinfo
  {journal} {Nonlinearity}\ }\textbf {\bibinfo {volume} {25}},\ \bibinfo
  {pages} {2245} (\bibinfo {year} {2012})}\BibitemShut {NoStop}%
\bibitem [{\citenamefont {Giomi}\ \emph {et~al.}(2013)\citenamefont {Giomi},
  \citenamefont {Bowick}, \citenamefont {Ma},\ and\ \citenamefont
  {Marchetti}}]{Giomi2013}%
  \BibitemOpen
  \bibfield  {author} {\bibinfo {author} {\bibfnamefont {L.}~\bibnamefont
  {Giomi}}, \bibinfo {author} {\bibfnamefont {M.~J.}\ \bibnamefont {Bowick}},
  \bibinfo {author} {\bibfnamefont {X.}~\bibnamefont {Ma}}, \ and\ \bibinfo
  {author} {\bibfnamefont {M.~C.}\ \bibnamefont {Marchetti}},\ }\href {\doibase
  10.1103/PhysRevLett.110.228101} {\bibfield  {journal} {\bibinfo  {journal}
  {Phys. Rev. Lett.}\ }\textbf {\bibinfo {volume} {110}},\ \bibinfo {pages}
  {228101} (\bibinfo {year} {2013})}\BibitemShut {NoStop}%
\bibitem [{\citenamefont {Thampi}\ \emph {et~al.}(2013)\citenamefont {Thampi},
  \citenamefont {Golestanian},\ and\ \citenamefont {Yeomans}}]{Thampi2013}%
  \BibitemOpen
  \bibfield  {author} {\bibinfo {author} {\bibfnamefont {S.~P.}\ \bibnamefont
  {Thampi}}, \bibinfo {author} {\bibfnamefont {R.}~\bibnamefont {Golestanian}},
  \ and\ \bibinfo {author} {\bibfnamefont {J.~M.}\ \bibnamefont {Yeomans}},\
  }\href {\doibase 10.1103/PhysRevLett.111.118101} {\bibfield  {journal}
  {\bibinfo  {journal} {Phys. Rev. Lett.}\ }\textbf {\bibinfo {volume} {111}},\
  \bibinfo {pages} {118101} (\bibinfo {year} {2013})}\BibitemShut {NoStop}%
\bibitem [{\citenamefont {Giomi}\ \emph {et~al.}(2014)\citenamefont {Giomi},
  \citenamefont {Bowick}, \citenamefont {Mishra}, \citenamefont {Sknepnek},\
  and\ \citenamefont {Cristina~Marchetti}}]{Giomi2014}%
  \BibitemOpen
  \bibfield  {author} {\bibinfo {author} {\bibfnamefont {L.}~\bibnamefont
  {Giomi}}, \bibinfo {author} {\bibfnamefont {M.~J.}\ \bibnamefont {Bowick}},
  \bibinfo {author} {\bibfnamefont {P.}~\bibnamefont {Mishra}}, \bibinfo
  {author} {\bibfnamefont {R.}~\bibnamefont {Sknepnek}}, \ and\ \bibinfo
  {author} {\bibfnamefont {M.}~\bibnamefont {Cristina~Marchetti}},\ }\href
  {\doibase 10.1098/rsta.2013.0365} {\bibfield  {journal} {\bibinfo  {journal}
  {Phil. Trans. R. Soc. Lond. A}\ }\textbf {\bibinfo {volume} {372}},\ \bibinfo
  {pages} {20130365} (\bibinfo {year} {2014})}\BibitemShut {NoStop}%
\bibitem [{\citenamefont {Tjhung}\ \emph {et~al.}(2015)\citenamefont {Tjhung},
  \citenamefont {Tiribocchi}, \citenamefont {Marenduzzo},\ and\ \citenamefont
  {Cates}}]{Tjhung2015}%
  \BibitemOpen
  \bibfield  {author} {\bibinfo {author} {\bibfnamefont {E.}~\bibnamefont
  {Tjhung}}, \bibinfo {author} {\bibfnamefont {A.}~\bibnamefont {Tiribocchi}},
  \bibinfo {author} {\bibfnamefont {D.}~\bibnamefont {Marenduzzo}}, \ and\
  \bibinfo {author} {\bibfnamefont {M.~E.}\ \bibnamefont {Cates}},\ }\href
  {\doibase 10.1038/ncomms6420} {\bibfield  {journal} {\bibinfo  {journal}
  {Nat. Commun.}\ }\textbf {\bibinfo {volume} {6}},\ \bibinfo {pages} {5420}
  (\bibinfo {year} {2015})}\BibitemShut {NoStop}%
\bibitem [{\citenamefont {Joanny}\ and\ \citenamefont
  {Ramaswamy}(2012)}]{Joanny2012}%
  \BibitemOpen
  \bibfield  {author} {\bibinfo {author} {\bibfnamefont {J.-F.}\ \bibnamefont
  {Joanny}}\ and\ \bibinfo {author} {\bibfnamefont {S.}~\bibnamefont
  {Ramaswamy}},\ }\href {\doibase 10.1017/jfm.2012.131} {\bibfield  {journal}
  {\bibinfo  {journal} {J. Fluid Mech.}\ }\textbf {\bibinfo {volume} {705}},\
  \bibinfo {pages} {46} (\bibinfo {year} {2012})}\BibitemShut {NoStop}%
\bibitem [{\citenamefont {Tjhung}\ \emph {et~al.}(2012)\citenamefont {Tjhung},
  \citenamefont {Marenduzzo},\ and\ \citenamefont {Cates}}]{Tjhung2012}%
  \BibitemOpen
  \bibfield  {author} {\bibinfo {author} {\bibfnamefont {E.}~\bibnamefont
  {Tjhung}}, \bibinfo {author} {\bibfnamefont {D.}~\bibnamefont {Marenduzzo}},
  \ and\ \bibinfo {author} {\bibfnamefont {M.~E.}\ \bibnamefont {Cates}},\
  }\href {\doibase 10.1073/pnas.1200843109} {\bibfield  {journal} {\bibinfo
  {journal} {Proc. Natl. Acad. Sci.}\ }\textbf {\bibinfo {volume} {109}},\
  \bibinfo {pages} {12381} (\bibinfo {year} {2012})}\BibitemShut {NoStop}%
\bibitem [{\citenamefont {Giomi}\ and\ \citenamefont
  {DeSimone}(2014)}]{GiomiDeSimone}%
  \BibitemOpen
  \bibfield  {author} {\bibinfo {author} {\bibfnamefont {L.}~\bibnamefont
  {Giomi}}\ and\ \bibinfo {author} {\bibfnamefont {A.}~\bibnamefont
  {DeSimone}},\ }\href {\doibase 10.1103/PhysRevLett.112.147802} {\bibfield
  {journal} {\bibinfo  {journal} {Phys. Rev. Lett.}\ }\textbf {\bibinfo
  {volume} {112}},\ \bibinfo {pages} {147802} (\bibinfo {year}
  {2014})}\BibitemShut {NoStop}%
\bibitem [{\citenamefont {Khoromskaia}\ and\ \citenamefont
  {Alexander}(2015)}]{Khoromskaia2015}%
  \BibitemOpen
  \bibfield  {author} {\bibinfo {author} {\bibfnamefont {D.}~\bibnamefont
  {Khoromskaia}}\ and\ \bibinfo {author} {\bibfnamefont {G.~P.}\ \bibnamefont
  {Alexander}},\ }\href {\doibase 10.1103/PhysRevE.92.062311} {\bibfield
  {journal} {\bibinfo  {journal} {Phys. Rev. E}\ }\textbf {\bibinfo {volume}
  {92}},\ \bibinfo {pages} {062311} (\bibinfo {year} {2015})}\BibitemShut
  {NoStop}%
\bibitem [{\citenamefont {Voituriez}\ \emph {et~al.}(2005)\citenamefont
  {Voituriez}, \citenamefont {Joanny},\ and\ \citenamefont
  {Prost}}]{Voituriez2005}%
  \BibitemOpen
  \bibfield  {author} {\bibinfo {author} {\bibfnamefont {R.}~\bibnamefont
  {Voituriez}}, \bibinfo {author} {\bibfnamefont {J.~F.}\ \bibnamefont
  {Joanny}}, \ and\ \bibinfo {author} {\bibfnamefont {J.}~\bibnamefont
  {Prost}},\ }\href {\doibase 10.1209/epl/i2004-10501-2} {\bibfield  {journal}
  {\bibinfo  {journal} {Europhys. Lett.}\ }\textbf {\bibinfo {volume} {70}},\
  \bibinfo {pages} {404} (\bibinfo {year} {2005})}\BibitemShut {NoStop}%
\bibitem [{\citenamefont {Sankararaman}\ and\ \citenamefont
  {Ramaswamy}(2009)}]{Sankararaman2009}%
  \BibitemOpen
  \bibfield  {author} {\bibinfo {author} {\bibfnamefont {S.}~\bibnamefont
  {Sankararaman}}\ and\ \bibinfo {author} {\bibfnamefont {S.}~\bibnamefont
  {Ramaswamy}},\ }\href {\doibase 10.1103/PhysRevLett.102.118107} {\bibfield
  {journal} {\bibinfo  {journal} {Phys. Rev. Lett.}\ }\textbf {\bibinfo
  {volume} {102}},\ \bibinfo {pages} {118107} (\bibinfo {year}
  {2009})}\BibitemShut {NoStop}%
\bibitem [{\citenamefont {Goldstein}\ \emph {et~al.}(2008)\citenamefont
  {Goldstein}, \citenamefont {Tuval},\ and\ \citenamefont {van~de
  Meent}}]{Goldstein2008}%
  \BibitemOpen
  \bibfield  {author} {\bibinfo {author} {\bibfnamefont {R.~E.}\ \bibnamefont
  {Goldstein}}, \bibinfo {author} {\bibfnamefont {I.}~\bibnamefont {Tuval}}, \
  and\ \bibinfo {author} {\bibfnamefont {J.-W.}\ \bibnamefont {van~de Meent}},\
  }\href {\doibase 10.1073/pnas.0707223105} {\bibfield  {journal} {\bibinfo
  {journal} {Proc. Natl. Acad. Sci.}\ }\textbf {\bibinfo {volume} {105}},\
  \bibinfo {pages} {3663} (\bibinfo {year} {2008})},\ \Eprint
  {http://arxiv.org/abs/http://www.pnas.org/content/105/10/3663.full.pdf}
  {http://www.pnas.org/content/105/10/3663.full.pdf} \BibitemShut {NoStop}%
\bibitem [{\citenamefont {Woodhouse}\ and\ \citenamefont
  {Goldstein}(2013)}]{Woodhouse2013}%
  \BibitemOpen
  \bibfield  {author} {\bibinfo {author} {\bibfnamefont {F.~G.}\ \bibnamefont
  {Woodhouse}}\ and\ \bibinfo {author} {\bibfnamefont {R.~E.}\ \bibnamefont
  {Goldstein}},\ }\href {\doibase 10.1073/pnas.1302736110} {\bibfield
  {journal} {\bibinfo  {journal} {Proc. Natl. Acad. Sci.}\ }\textbf {\bibinfo
  {volume} {110}},\ \bibinfo {pages} {14132} (\bibinfo {year} {2013})},\
  \Eprint
  {http://arxiv.org/abs/http://www.pnas.org/content/110/35/14132.full.pdf}
  {http://www.pnas.org/content/110/35/14132.full.pdf} \BibitemShut {NoStop}%
\bibitem [{\citenamefont {Doostmohammadi}\ \emph {et~al.}(2015)\citenamefont
  {Doostmohammadi}, \citenamefont {Thampi}, \citenamefont {Saw}, \citenamefont
  {Lim}, \citenamefont {Ladoux},\ and\ \citenamefont
  {Yeomans}}]{Doostmohammadi2015}%
  \BibitemOpen
  \bibfield  {author} {\bibinfo {author} {\bibfnamefont {A.}~\bibnamefont
  {Doostmohammadi}}, \bibinfo {author} {\bibfnamefont {S.~P.}\ \bibnamefont
  {Thampi}}, \bibinfo {author} {\bibfnamefont {T.~B.}\ \bibnamefont {Saw}},
  \bibinfo {author} {\bibfnamefont {C.~T.}\ \bibnamefont {Lim}}, \bibinfo
  {author} {\bibfnamefont {B.}~\bibnamefont {Ladoux}}, \ and\ \bibinfo {author}
  {\bibfnamefont {J.~M.}\ \bibnamefont {Yeomans}},\ }\href {\doibase
  10.1039/C5SM01382H} {\bibfield  {journal} {\bibinfo  {journal} {Soft Matter}\
  }\textbf {\bibinfo {volume} {11}},\ \bibinfo {pages} {7328} (\bibinfo {year}
  {2015})}\BibitemShut {NoStop}%
\bibitem [{\citenamefont {Kirby}(2010)}]{KirbyBook}%
  \BibitemOpen
  \bibfield  {author} {\bibinfo {author} {\bibfnamefont {B.~J.}\ \bibnamefont
  {Kirby}},\ }\href@noop {} {\emph {\bibinfo {title} {Micro and Nanoscale Fluid
  Mechanics: Transport in Microfluidic Devices}}}\ (\bibinfo  {publisher}
  {Cambridge University Press},\ \bibinfo {year} {2010})\BibitemShut {NoStop}%
\bibitem [{\citenamefont {Marenduzzo}\ \emph {et~al.}(2007)\citenamefont
  {Marenduzzo}, \citenamefont {Orlandini}, \citenamefont {Cates},\ and\
  \citenamefont {Yeomans}}]{Davide2007}%
  \BibitemOpen
  \bibfield  {author} {\bibinfo {author} {\bibfnamefont {D.}~\bibnamefont
  {Marenduzzo}}, \bibinfo {author} {\bibfnamefont {E.}~\bibnamefont
  {Orlandini}}, \bibinfo {author} {\bibfnamefont {M.~E.}\ \bibnamefont
  {Cates}}, \ and\ \bibinfo {author} {\bibfnamefont {J.~M.}\ \bibnamefont
  {Yeomans}},\ }\href {\doibase 10.1103/PhysRevE.76.031921} {\bibfield
  {journal} {\bibinfo  {journal} {Phys. Rev. E}\ }\textbf {\bibinfo {volume}
  {76}},\ \bibinfo {pages} {031921} (\bibinfo {year} {2007})}\BibitemShut
  {NoStop}%
\bibitem [{\citenamefont {Blow}\ \emph {et~al.}(2014)\citenamefont {Blow},
  \citenamefont {Thampi},\ and\ \citenamefont {Yeomans}}]{Blow2014}%
  \BibitemOpen
  \bibfield  {author} {\bibinfo {author} {\bibfnamefont {M.~L.}\ \bibnamefont
  {Blow}}, \bibinfo {author} {\bibfnamefont {S.~P.}\ \bibnamefont {Thampi}}, \
  and\ \bibinfo {author} {\bibfnamefont {J.~M.}\ \bibnamefont {Yeomans}},\
  }\href {\doibase 10.1103/PhysRevLett.113.248303} {\bibfield  {journal}
  {\bibinfo  {journal} {Phys. Rev. Lett.}\ }\textbf {\bibinfo {volume} {113}},\
  \bibinfo {pages} {248303} (\bibinfo {year} {2014})}\BibitemShut {NoStop}%
\bibitem [{\citenamefont {Love}(1927)}]{LoveBook}%
  \BibitemOpen
  \bibfield  {author} {\bibinfo {author} {\bibfnamefont {A.~E.~H.}\
  \bibnamefont {Love}},\ }\href@noop {} {\emph {\bibinfo {title} {A Treatise on
  the Mathematical Theory of Elasticity}}}\ (\bibinfo  {publisher} {Cambridge
  University Press},\ \bibinfo {year} {1927})\BibitemShut {NoStop}%
\bibitem [{\citenamefont {de~Gennes}\ and\ \citenamefont
  {Prost}(1995)}]{DeGennesBook}%
  \BibitemOpen
  \bibfield  {author} {\bibinfo {author} {\bibfnamefont {P.~G.}\ \bibnamefont
  {de~Gennes}}\ and\ \bibinfo {author} {\bibfnamefont {J.}~\bibnamefont
  {Prost}},\ }\href@noop {} {\emph {\bibinfo {title} {The Physics of Liquid
  Crystals}}}\ (\bibinfo  {publisher} {Oxford University Press},\ \bibinfo
  {year} {1995})\BibitemShut {NoStop}%
\bibitem [{\citenamefont {Chaikin}\ and\ \citenamefont
  {Lubensky}(2000)}]{ChaikinBook}%
  \BibitemOpen
  \bibfield  {author} {\bibinfo {author} {\bibfnamefont {P.}~\bibnamefont
  {Chaikin}}\ and\ \bibinfo {author} {\bibfnamefont {T.}~\bibnamefont
  {Lubensky}},\ }\href@noop {} {\emph {\bibinfo {title} {Principles of
  Condensed Matter Physics}}}\ (\bibinfo  {publisher} {Cambridge University
  Press},\ \bibinfo {year} {2000})\BibitemShut {NoStop}%
\bibitem [{\citenamefont {Orlandini}\ \emph {et~al.}(1995)\citenamefont
  {Orlandini}, \citenamefont {Swift},\ and\ \citenamefont
  {Yeomans}}]{Orlandini1995}%
  \BibitemOpen
  \bibfield  {author} {\bibinfo {author} {\bibfnamefont {E.}~\bibnamefont
  {Orlandini}}, \bibinfo {author} {\bibfnamefont {M.~R.}\ \bibnamefont
  {Swift}}, \ and\ \bibinfo {author} {\bibfnamefont {J.~M.}\ \bibnamefont
  {Yeomans}},\ }\href {\doibase 10.1209/0295-5075/32/6/001} {\bibfield
  {journal} {\bibinfo  {journal} {Europhys. Lett.}\ }\textbf {\bibinfo {volume}
  {32}},\ \bibinfo {pages} {463} (\bibinfo {year} {1995})}\BibitemShut
  {NoStop}%
\bibitem [{\citenamefont {Sulaiman}\ \emph {et~al.}(2006)\citenamefont
  {Sulaiman}, \citenamefont {Marenduzzo},\ and\ \citenamefont
  {Yeomans}}]{Sulaiman2006}%
  \BibitemOpen
  \bibfield  {author} {\bibinfo {author} {\bibfnamefont {N.}~\bibnamefont
  {Sulaiman}}, \bibinfo {author} {\bibfnamefont {D.}~\bibnamefont
  {Marenduzzo}}, \ and\ \bibinfo {author} {\bibfnamefont {J.~M.}\ \bibnamefont
  {Yeomans}},\ }\href {\doibase 10.1103/PhysRevE.74.041708} {\bibfield
  {journal} {\bibinfo  {journal} {Phys. Rev. E}\ }\textbf {\bibinfo {volume}
  {74}},\ \bibinfo {pages} {041708} (\bibinfo {year} {2006})}\BibitemShut
  {NoStop}%
\bibitem [{\citenamefont {Cahn}\ and\ \citenamefont
  {Hilliard}(1958)}]{Cahn1958}%
  \BibitemOpen
  \bibfield  {author} {\bibinfo {author} {\bibfnamefont {J.~W.}\ \bibnamefont
  {Cahn}}\ and\ \bibinfo {author} {\bibfnamefont {J.~E.}\ \bibnamefont
  {Hilliard}},\ }\href {\doibase 10.1063/1.1744102} {\bibfield  {journal}
  {\bibinfo  {journal} {J. Chem. Phys.}\ }\textbf {\bibinfo {volume} {28}},\
  \bibinfo {pages} {258} (\bibinfo {year} {1958})}\BibitemShut {NoStop}%
\bibitem [{\citenamefont {Beris}\ and\ \citenamefont
  {Edwards}(1994)}]{BerisBook}%
  \BibitemOpen
  \bibfield  {author} {\bibinfo {author} {\bibfnamefont {A.~N.}\ \bibnamefont
  {Beris}}\ and\ \bibinfo {author} {\bibfnamefont {B.~J.}\ \bibnamefont
  {Edwards}},\ }\href@noop {} {\emph {\bibinfo {title} {Thermodynamics of
  Flowing Systems}}}\ (\bibinfo  {publisher} {Oxford University Press},\
  \bibinfo {year} {1994})\BibitemShut {NoStop}%
\bibitem [{\citenamefont {A.~Simha}\ and\ \citenamefont
  {Ramaswamy}(2002)}]{Simha2002}%
  \BibitemOpen
  \bibfield  {author} {\bibinfo {author} {\bibfnamefont {R.}~\bibnamefont
  {A.~Simha}}\ and\ \bibinfo {author} {\bibfnamefont {S.}~\bibnamefont
  {Ramaswamy}},\ }\href {\doibase 10.1103/PhysRevLett.89.058101} {\bibfield
  {journal} {\bibinfo  {journal} {Phys. Rev. Lett.}\ }\textbf {\bibinfo
  {volume} {89}},\ \bibinfo {pages} {058101} (\bibinfo {year}
  {2002})}\BibitemShut {NoStop}%
\bibitem [{\citenamefont {Ramaswamy}(2010)}]{Ramaswamy2010}%
  \BibitemOpen
  \bibfield  {author} {\bibinfo {author} {\bibfnamefont {S.}~\bibnamefont
  {Ramaswamy}},\ }\href {\doibase 10.1146/annurev-conmatphys-070909-104101}
  {\bibfield  {journal} {\bibinfo  {journal} {Annu. Rev. Cond. Mat. Phys.}\
  }\textbf {\bibinfo {volume} {1}},\ \bibinfo {pages} {323} (\bibinfo {year}
  {2010})}\BibitemShut {NoStop}%
\bibitem [{\citenamefont {Grzybowski}\ \emph {et~al.}(2001)\citenamefont
  {Grzybowski}, \citenamefont {Bowden}, \citenamefont {Arias}, \citenamefont
  {Yang},\ and\ \citenamefont {Whitesides}}]{Grzybowski2001}%
  \BibitemOpen
  \bibfield  {author} {\bibinfo {author} {\bibfnamefont {B.~A.}\ \bibnamefont
  {Grzybowski}}, \bibinfo {author} {\bibfnamefont {N.}~\bibnamefont {Bowden}},
  \bibinfo {author} {\bibfnamefont {F.}~\bibnamefont {Arias}}, \bibinfo
  {author} {\bibfnamefont {H.}~\bibnamefont {Yang}}, \ and\ \bibinfo {author}
  {\bibfnamefont {G.~M.}\ \bibnamefont {Whitesides}},\ }\href {\doibase
  10.1021/jp0026383} {\bibfield  {journal} {\bibinfo  {journal} {J. Phys. Chem.
  B}\ }\textbf {\bibinfo {volume} {105}},\ \bibinfo {pages} {404} (\bibinfo
  {year} {2001})}\BibitemShut {NoStop}%
\bibitem [{\citenamefont {Brown}\ \emph {et~al.}(2009)\citenamefont {Brown},
  \citenamefont {Wells}, \citenamefont {Newton},\ and\ \citenamefont
  {McHale}}]{Brown2009}%
  \BibitemOpen
  \bibfield  {author} {\bibinfo {author} {\bibfnamefont {C.~V.}\ \bibnamefont
  {Brown}}, \bibinfo {author} {\bibfnamefont {G.~G.}\ \bibnamefont {Wells}},
  \bibinfo {author} {\bibfnamefont {M.~I.}\ \bibnamefont {Newton}}, \ and\
  \bibinfo {author} {\bibfnamefont {G.}~\bibnamefont {McHale}},\ }\href
  {http://dx.doi.org/10.1038/nphoton.2009.99} {\bibfield  {journal} {\bibinfo
  {journal} {Nat. Photonics.}\ }\textbf {\bibinfo {volume} {3}},\ \bibinfo
  {pages} {403} (\bibinfo {year} {2009})}\BibitemShut {NoStop}%
\end{thebibliography}%

\end{document}